\documentclass[aps,prb,twocolumn,superscriptaddress, show pacs]{revtex4-1}

\usepackage{bm, amsmath}
\usepackage{bm}
\usepackage[euler]{textgreek}

\usepackage{graphicx}
\usepackage{subfigure}
\usepackage[outdir=./]{epstopdf}

\usepackage{dcolumn}

\usepackage[normalem]{ulem} 
\usepackage{verbatim}
\usepackage[draft]{todonotes} 
\usepackage{float}

\newcommand{\V}[1]{\mathbf{#1}} 
\newcommand{\U}[1]{\hat{\mathbf{#1}}} 

\begin{document}

\title{Neutron study of in-plane skyrmions in MnSi thin films}

\author{S.~A.~Meynell}
\author{M.~N.~Wilson}
\affiliation{Department of Physics and Atmospheric Science, Dalhousie University, Halifax, Nova Scotia, Canada B3H 4R2}

\author{K.~L.~Krycka}
\author{B.~J.~Kirby}
\affiliation{Center for Neutron Research, NIST, Gaithersburg Maryland 20899, USA}

\author{H.~Fritzsche}
\affiliation{Canadian Nuclear Laboratories, Chalk River, Ontario, Canada K0J 1J0}

\author{T.~L.~Monchesky} \thanks{tmonches@dal.ca}
\affiliation{Department of Physics and Atmospheric Science, Dalhousie University, Halifax, Nova Scotia, Canada B3H 3J5}

\date{\today}

\begin{abstract}
{ 
The magnetic structure of the in-plane skyrmions in epitaxial MnSi/Si(111) thin films is probed in three dimensions by the combination of polarized neutron reflectometry (PNR) and small angle neutron scattering (SANS).  We demonstrate that skyrmions exist in a region of the phase diagram above at temperature of 10 K.  PNR shows the skyrmions are confined to the middle of the film due to the potential well formed by the surface twists.  However, SANS shows that there is considerable disorder within the plane indicating that the magnetic structure is a 2D skyrmion glass.  
}
\end{abstract}


\maketitle

\section{Introduction}
The chiral magnetic interaction discovered by Dzyaloshinskii \cite{Dzyaloshinskii:1958jpcs} and Moriya \cite{Moriya:1960pr} plays an important role in the formation of magnetic nanostructures lacking inversion symmetry. 
The broken symmetry brought on by the presence of interfaces \cite{Fert:1990msf} or a chirality in the bulk crystal structure \cite{Dzyaloshinskii:1964zetf-2, Moriya:1960pr}  leads to a twisting of the magnetic order.  
Bogdanov first identified this Dzyaloshinskii-Moriya (DM) interaction's unusual ability to create stable cylindrical magnetic solitons, known as \emph{skyrmions} \cite{Bogdanov:1989jetp,Bogdanov:1994jmmm}, that could potentially form the basis for low-energy magnetic storage devices \cite{Kiselev:2011jpd,Sampaio:2013nn, Nagaosa:2013nn}.  
While a chiral interaction is a necessary ingredient  to stabilize multi-dimensional solitons in magnetic materials \cite{Bogdanov:1989jetp,Bogdanov:1994jmmm} and other condensed matter systems \cite{Bogdanov:2000jetp, Leonov:2014pre, Ackerman:2014pre}, additional interactions are required to make the skyrmions thermodynamically stable.  
The importance of various contributions to their stability remains an active area of investigation in both bulk crystals\cite{Rossler:2006nat,Muhlbauer:2009sci,Pappas:2009prl,Wilhelm:2011prl} and in nanostructures.

In the case of magnetic nanostructures, anisotropy \cite{Butenko:2010prb, Vousden:2016apl} and finite size effects \cite{Rybakov:2013prb, Rybakov:2015prl} are two key mechanisms that give the skyrmion phase its robustness over a large temperature and field range.  The latter likely plays the dominant role in creating stable skyrmions in free-standing chiral magnetic nano-crystals \cite{Yu:2010nat, Yu:2011nm,Tonomura:2012nl,Du:2015nc2}.  Studies of the thickness dependence of the magnetic structure in wedge-shaped specimens reveals the influence of confinement on the skyrmions' stability \cite{Yu:2015prb, Leonov:2016prl}. 
\begin{figure}[!h]
	\centering
	\includegraphics[width = \columnwidth]{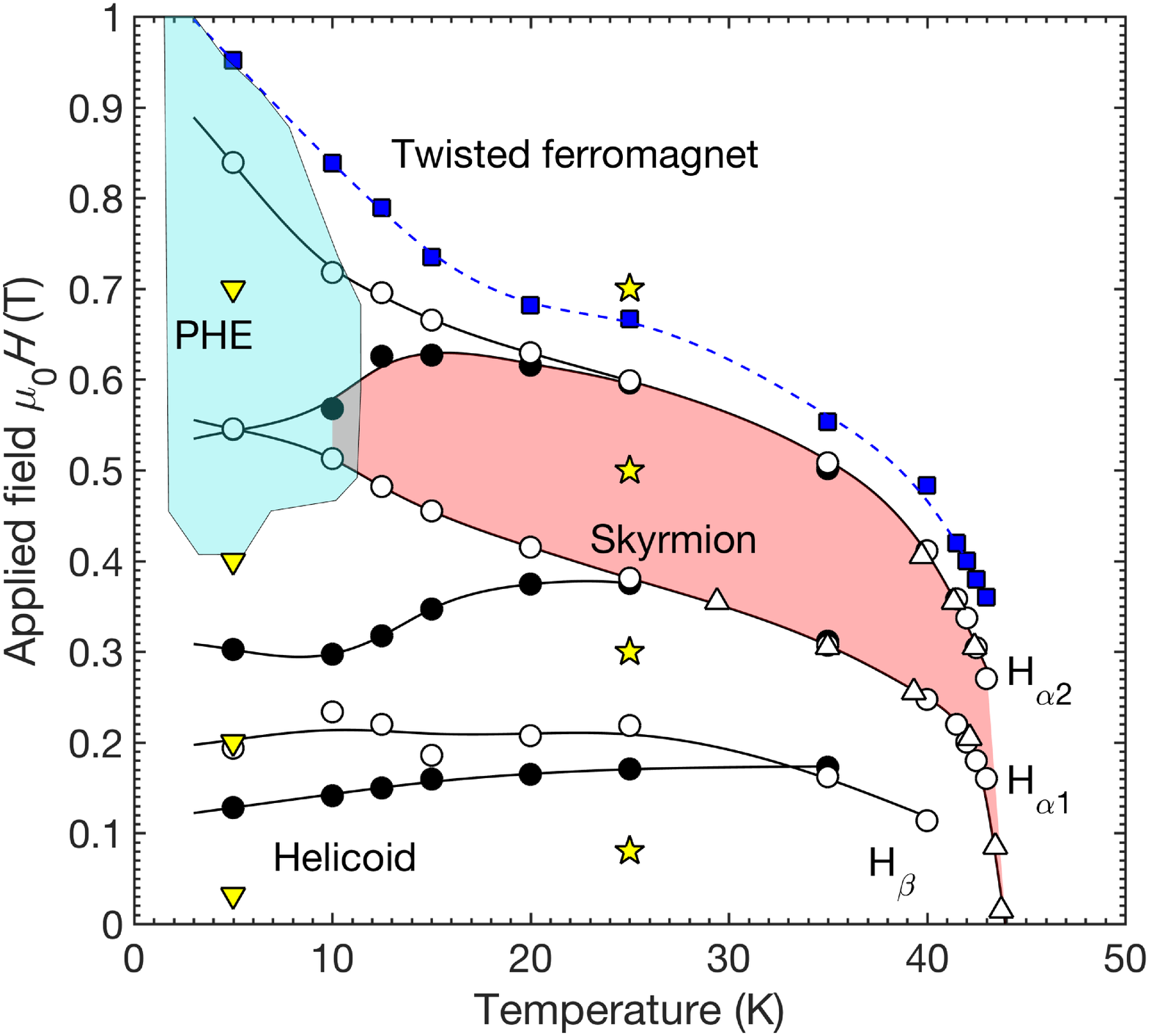}
	\caption{ (Color online) The phase diagram for a MnSi/Si(111) film with $\mathbf{H}\|[1\bar{1}0]$.   The filled (unfilled) circles correspond to peaks in $dM/dH$ measured in decreasing (increasing) magnetic fields for a $d = 26.7$~nm sample.  The blue squares are minima in $dM^2/dH^2$ for increasing magnetic fields \cite{Wilson:2012prb}.  The PNR experiments were performed in successively decreasing magnetic fields at the $(H,T)$ points shown by the yellow stars (this study), and by the yellow triangles (Ref.~\onlinecite{Wilson:2013prb}).  The cyan-colored area taken from Ref.~\onlinecite{Tomoyuki:2015jpsj} shows the region where a drop in the planar Hall effect signal is observed in a similar $d = 26$~nm sample measured in a decreasing field.}
	\label{fig:PD}
\end{figure}

Anisotropy is of importance for understanding the magnetic structure of chiral magnetic epilayers grown on Si(111) substrates, including MnSi \cite{Karhu:2010prb,Karhu:2011prb, Karhu:2012prb}, FeGe \cite{Huang:2012prl}, Fe$_x$Co$_{1-x}$Si \cite{Porter:2012prb, Sinha:2014prb} and MnGe \cite{Engelke:2012jpsp}.  A hard-axis out-of-plane magnetocrystalline anisotropy exists in the case of MnSi/Si(111) \cite{Karhu:2012prb}.  Although numerical simulations show that finite size effects can lead to stable skyrmion states for small enough values of the hard-axis anisotropy and film thickness \cite{Vousden:2016apl}, no evidence for skyrmions are found in magnetometry or electron transport measurements in MnSi/Si(111) for film thicknesses greater than 10~nm for out-of-plane fields  \cite{Wilson:2014prb}.  A muon-spin rotation study hints at possible additional magnetic phases \cite{Lancaster:2016prb}, and a combined Lorentz microscopy and Hall effect study claims to find evidence for in-plane helicoids and out-of-plane skyrmions\cite{Li:2013prl}. However, these microscopy results can be explained by structural artifacts \cite{Monchesky:2014prl} and the small anomalous Hall effect is explained by non-adiabatic spin transport in the conical phase without a topological contribution from a skyrmion phase \cite{Meynell:2014prb2}.

Application of a magnetic field in the easy-plane of MnSi/Si(111) changes the situation.  Even in the absence of finite size effects, micromagnetic calculations show that the magnetocrystalline anisotropy is capable of producing stable skyrmion lattices for a range of magnetic field strengths \cite{Wilson:2012prb}.  SQUID magnetometry reveals a set of first-order magnetic phase transitions that are entirely absent in out-of-plane fields for thicknesses between 0.8 and 3.5 times the zero-field helical wavelength, $L_D = 13.9$~nm.  In Ref.~\onlinecite{Wilson:2012prb}, we constructed the magnetic phase diagram for a $d = 1.9 L_D$ MnSi film in in-plane magnetic fields from the static susceptibility, $dM/dH$.  First-order magnetic phase transitions were identified by peaks in the susceptibility.  Such peaks are due to energy barriers that arise, for instance, in the transition between magnetic states with different topology.  However, the interpretation of these transitions, reproduced in Fig.~\ref{fig:PD}, is controversial.  At low temperatures, below approximately $T \simeq 12$~K, two peaks are found in the field dependence of $dM/dH$.  PNR in concert with magnetometry measurements demonstrate that these peaks mark the discrete unwinding of the helicoid from a two-turn to a one-turn and finally to a twisted ferromagnetic state. 

Conversely, Ref.~\onlinecite{Tomoyuki:2015jpsj} claims to find indirect evidence for skyrmions from planar Hall effect (PHE) measurements in this same low temperature region of the phase diagram. However, the author's interpretation contradicts the earlier PNR measurements that provide a direct measure of the magnetic structure.   Reference \onlinecite{Tomoyuki:2015jpsj} interprets a drop in the PHE resistivity at low $T$ and high $H$ observed in the cyan colored region in Fig.~\ref{fig:PD} as evidence of an in-plane skyrmion phase. We, however, claim the in-plane skyrmion phase is only observed in the red shaded region of Fig.~\ref{fig:PD}.  In this region -- for temperatures $12 \lesssim T \lesssim 42 $~K -- three critical fields, denoted as $H_{\alpha 2}$, $H_{\alpha 1}$, and $H_{\beta}$, provide evidence for the appearance of a magnetic phase not observed at low temperature. This phase is purportedly a distinct class of skyrmions with their core magnetizations pointing in the plane of the film \cite{Wilson:2012prb}.  To address the disagreement over the phase diagram, we report on a study of the magnetic structure of MnSi/Si(111) with a combination of polarized neutron reflectometry (PNR) and small angle neutron scattering (SANS).  

\section{Model}
We modeled the magnetic states of MnSi thin films explored in the neutron scattering experiments with the Bak Jensen free energy density,\cite{Bak:1980jpc} $w$,%
\begin{equation}
	w = A(\nabla \U{m})^2 + D \U{m} \cdot \nabla \times \U{m} -
	K (\U{m} \cdot \U{n})^2 - \mu_0 (\V{H} +	\frac{1}{2} \V{H}_d) \cdot \V{M},
	\label{eq:Bak}
\end{equation}
which we have used previously in Ref.~\onlinecite{Wilson:2012prb}.  Here, $\U{m}$ is a unit vector along the direction of the magnetization $\V{M}(\V{r})$ and $H$ is the magnetic field. The competition between exchange and the DM interaction, parameterized by $A$ and $D$ respectively, sets the helical wavelength, $L_D = 4\pi A/D$. The model only includes a uniaxial magnetocrystalline anisotropy, $K$, due to epitaxial strain. In the case of MnSi, the Si substrate induces a compression of the (111) planes and results in $K < 0$. \cite{Karhu:2012prb}  Solutions to Eq.~\ref{eq:Bak} for out-of-plane fields in the film limit give rise to simple 1-dimensional (1D) solutions, namely helical and conical states\cite{Wilson:2014prb}.  The intrinsic transition field between the conical and saturated states is $\mu_0 H_D = D^2/(2AM_s)$, where $M_s$ is the saturation magnetization. In the case of a helical film, this is modified by the demagnetizing field, $\V{H}_d = - \V{M} \cdot \hat{\V{n}} $ and the effective field due to the uniaxial anisotropy, $\mu_0 H_u = 2 K /M_s$, to produce the out-of-plane saturation field\cite{Karhu:2012prb}:
\begin{equation}
	H_{sat}^{\perp} = H_D - H_u + M_s.
	\label{eq:sat}
\end{equation}
For in-plane magnetic fields, the solutions to Eq.~(\ref{eq:Bak}) are more complex and lead to helicoids \cite{Wilson:2013prb} and skyrmions \cite{Wilson:2012prb}.  These magnetic textures were explored by finite difference methods with a steepest-descent solver implemented in MuMax3 \cite{Vansteenkiste:2014a}.  The magnetization was calculated on a 2D grid with 256 cells along the z-axis, and 4096 along the y-direction, with square or nearly square cell dimensions.  Periodic boundary conditions are used along the in-plane $x$- and $y$-directions.  The demagnetizing field is neglected in the calculation since it plays a minor role relative to the DM interaction \cite{Kiselev:2011jpd}.

\section{Experiment}
While Lorentz microscopy has been successfully used to study out-of-plane skyrmions \cite{Yu:2010nat, Yu:2011nm,Tonomura:2012nl}, the in-plane skyrmion lattices are generally not amenable to this technique, which is sensitive to the local magnetization averaged over the film thickness. For TEM measurements in a cross-sectional geometry, the Fresnel fringes generated by the film/substrate and film/cap interfaces wash out any magnetic contrast, unless the electron scattering potential of the chiral magnet is matched to the surrounding material, as was accomplished in Pt/FeGe/Pt nanostructures \cite{Du:2015nc2}. This is not the case for MnSi/Si interfaces where there is appreciable electron density differences, and neutron scattering is better suited to investigate the buried magnetic structures in this case.
   
In this paper, we focus on MnSi films $L_D < d < 2 L_D$, where only one row of skrymions are theoretically expected to form in the middle of the film \cite{Wilson:2012prb}.  MnSi films with thicknesses of either $d = 25.2$~nm or 26.7~nm were grown by molecular beam epitaxy on Si(111) substrates.  The films were capped with 20 nm of amorphous Si to protect them from oxidation.   The films' crystal structures were confirmed with X-ray diffraction and X-ray reflectometry to screen for any secondary phase.   Details about the sample preparation and growth are given in Ref.~\onlinecite{Karhu:2012prb, Meynell:2014prb2}.
   
\begin{figure}
\includegraphics[width=0.8\columnwidth]{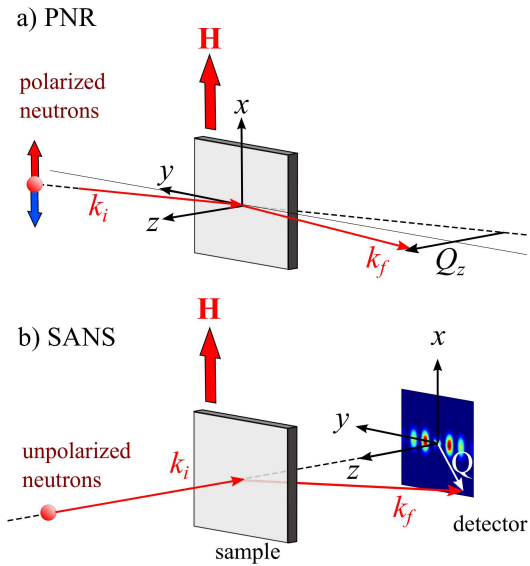}
\caption{
(Color online) Experimental geometries used for a) polarized neutron reflectometry, and b) SANS.
}
\label{fig:neutron}
\end{figure}

We compare the two geometries used in the two neutron scattering experiments in Fig.~\ref{fig:neutron}.   In the case of PNR, the specularly reflected beam is measured as a function of angle of incidence such that the scattering vector of the measured intensity is always perpendicular to the film surface.   These experiments are only sensitive to the depth profile of the nuclear and the magnetic scattering length densities where in-plane variations are averaged out.  The magnetic component of the non-spin flip reflectivities is determined by the component of the magnetization along the field direction, here taken to be the $x$-direction (Fig.~\ref{fig:neutron}(a)).

In contrast to PNR, the incident neutron beam in the SANS experiment is along the film normal ($z$-direction), as illustrated Fig.~\ref{fig:neutron}(b), and the transmitted beam is measured with a 2D detector such that the measured intensity is from scattering vectors $\mathbf{Q}$ that lie in the plane of the film.  This provides a measure of the in-plane variation of the magnetization while averaging over variations across the depth of the film.  These complementary geometries enable us to probe the magnetic structure in all three dimensions.

\subsection{PNR}

PNR measurements were conducted at the Canadian Nuclear Laboratories with 0.237~nm neutrons on the D3 reflectometer with an electromagnet and closed-cycle cryostat \cite{Fritzsche:2005rsi}.  A spin polarization in excess of 95\% was produced by a CoFe/Si supermirror and a Mezei-type precession spin flipper.  Previous measurements showed that the spin-flip signal is negligible due to the cancelling effects of the bi-chiral magnetic domains \cite{Karhu:2012prb}.  Therefore we measured the spin-up, $R_+$ and spin-down, $R_-$ reflectivities without an analyzer in order to increase the signal-to-noise ratio.  Simulations performed using SimulReflec software included a correction for the flipping ratio of the polarizer (see Ref.~\onlinecite{Meynell:2014prb1}).  The magnetic field $\mu_0 H$ was applied along the in-plane MnSi$[1\bar{1}0]$ direction, denoted as the $x$ direction in Fig.~\ref{fig:neutron} (a). 

\begin{figure}
\centering
\includegraphics[width=\columnwidth]{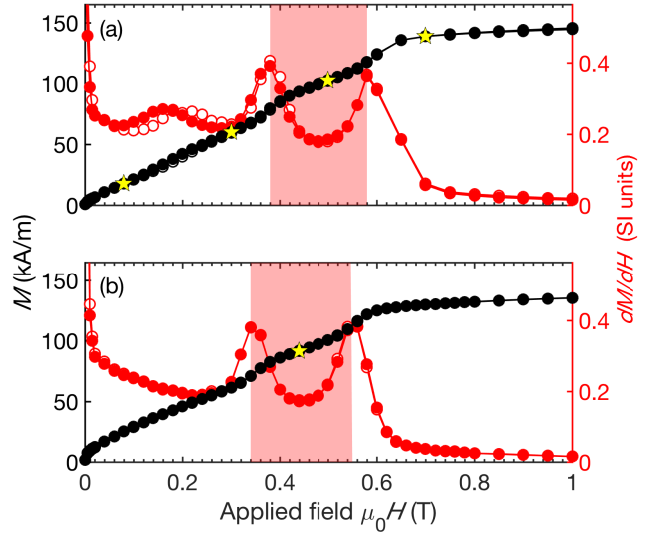}
\caption{
(Color online) Magnetization (in black) and the static susceptibility $dM/dH$ is shown (in red) of MnSi films for magnetic fields $\mathbf{H}||[1\bar{1}0]$ at $T=25$~K measured by SQUID magnetometry. The unfilled circles are measurements in increasing applied field $H$, and the filled circles are in decreasing $H$. (a) $d=26.7$~nm thick MnSi layer, and (b) $d=25.4$~nm thick layer.  The stars denote the field value of the neutron experiments presented in Figs.~\ref{fig:PNR} and \ref{fig:SANS}.
	}
\label{fig:hys}
\end{figure} 

The stars in Fig.~\ref{fig:PD} show the $(H,T)$ points in the phase diagram sampled by the PNR measurements. The magnetic state of the sample was prepared by cooling in a magnetic field $\mu_0 H = 0.8$~T from $T=100$~K to $T=25$~K, where a series of PNR measurements were performed in consecutively decreasing magnetic fields across the four regions of the magnetic phase diagram.  At  $T=25$~K, the first-order magnetic phase transitions occur at fields $\mu_0 H_{\alpha 2} = 0.60$~T, $\mu_0 H_{\alpha 1} = 0.38$~T, and $\mu_0 H_{\beta} = 0.22$~T.  We chose fields in between these critical fields and avoided fields under the peaks in $dM/dH$ in Fig.~\ref{fig:hys} since first-order transitions are characterized by regions of mixed phases that would complicate the analysis.  

An accurate measurement of $H_D$ is more difficult in thin films, but needed for a comparison with theory.  To estimate $H_D$ we analyzed $M$-$H$ loops measured by SQUID magnetometry.  For in-plane fields, the surface twists that exist in the field induced ferromagnetic state never fully saturate and complicate the analysis.  These surface twists largely disappear for out-of-plane magnetic fields, and the saturation field is given by Eq.~\ref{eq:sat}. While $H_D$ cannot be determined from the out-of-plane measurement alone, it is possible to obtain the sum of $H_D$ and $H_u$:  from Fig.~\ref{fig:hys} we find $\mu_0(H_D - H_u) = 0.86$~T for the $d=26.7$~nm sample at $T=25$~K, (obtained from $\mu_0 H_{sat}^{\perp} = 1.03$~T and $M_s$ = 138 kA/m).  To separate the DM interaction from the anisotropy, we note that the penetration depth of the surface twists, measured by PNR, is determined by $H_D$ and $L_D$ (see Eqs.~(9-11) in Ref.~\onlinecite{Meynell:2014prb1}). We therefore treat $H_D$ as a fitting parameter in the analysis of the high-field PNR data below.  
 
\begin{figure}
\centering
\includegraphics[width=\columnwidth]{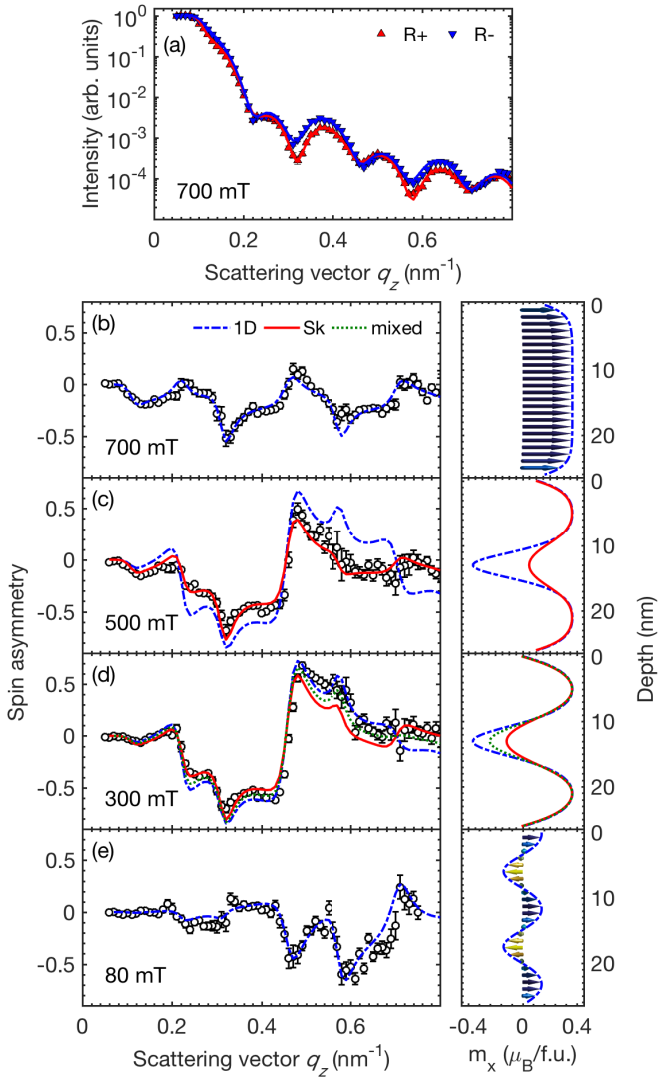}
\caption{
(Color online) PNR of a $d=26.7$~nm MnSi film for magnetic fields $\mathbf{H}||[1\bar{1}0]$ at $T=25$~K. (a) Polarized reflectivities at $\mu_0 H = 0.7$~T in the twisted magnetic state and the corresponding spin asymmetry in (b).  The points with $\pm \sigma$ error bars in (b) - (e) show the spin asymmetry in fields  $\mu_0 H = 0.7, 0.5, 0.3$ and 0.08~T as indicated, and the solid and dashed lines show simulations from various models as described in the text.  The insets show the magnetization depth profile used in each of the models. The error bars are $\pm 1 \sigma$.
}
\label{fig:PNR}
\end{figure} 

In Fig.~\ref{fig:PNR}(a) we show the reflectivity measurements, $R_{+}$ and $R_{-}$, in this case in a field of $\mu_0 H = 0.7$~T, together with the numerical calculations shown by the solid, dashed or dotted lines. The corresponding depth profile of the $x$-component of the magnetic moment $m$ is shown in the right-hand panel of Fig.~\ref{fig:PNR}(b).  The simulations include the nuclear scattering length densities obtained from combined x-ray reflectometry and PNR measurements \cite{Karhu:2012prb}.  To make the comparison with model calculations more clear, the $\mu_0 H = 0.7$~T data are presented in terms of the spin asymmetry, $\alpha = (R_{+} - R_{-})/ (R_{+} + R_{-})$, in the left-hand panel of Fig.~\ref{fig:PNR}(b).  For the three remaining field values, the model calculations fit $\alpha$, $R_{+}$ and $R_{-}$ equally well and therefore only $\alpha$ is shown.    Similar to measurements at $T=5$~K,\cite{Wilson:2013prb} we find that the magnetic structure is in a twisted ferromagnetic state in fields above $H_{\alpha 2}$.  By using the experimentally determined values $L_D$ and $M_s$ as input parameters, we calculated the magnetic structure using the discrete helicoidal model of Ref.~\onlinecite{Wilson:2013prb} with $H_D$ as the sole fitting parameter.   A fit to $\alpha$ in Fig.~\ref{fig:PNR}(b) yields $\mu_0 H_D = 1.06 \pm 0.26$~T. The absence of evidence of skyrmions in out-of-plane magnetic fields constrains $H_u < 0$, which leads to a best estimate for $\mu_0 H_D$ of approximately 0.85~T. The estimate for $\mu_0 H_D =0.85~T$ is supported by the lower field data as higher values for $H_D$ lead to worse fits for those fields. 

\begin{figure}
\includegraphics[width=1.0\columnwidth]{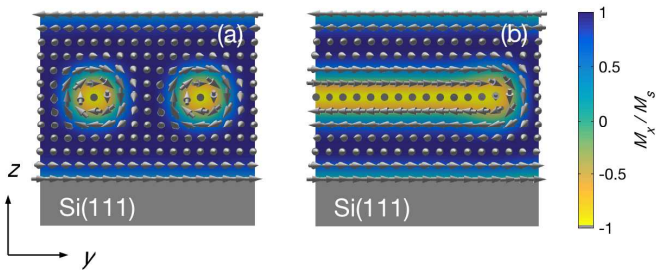}
\caption{
(Color online) The calculation of the magnetization $\V{M}$ in the $y-z$ plane of a 26.4 nm thick MnSi films, where the color-plot corresponds to $M_x$. (a)  Skyrmion grating for a field of $\mu_0 H = 0.5$~T. (b) Half of a metastable helicoidal structure at a field of $\mu_0 H = 0.3$~T, formed from a skyrmion when the applied field is dropped below the skyrmion elliptic instability. 
}
\label{fig:mumax}
\end{figure}

The set of measurements in Fig.~\ref{fig:PNR}(c) for $\mu_0 H = 0.5$~T represents the region in the phase diagram between $H_{\alpha 2}$ and $H_{\alpha 1}$.  Unlike the measurements at $T=5K$ for the same field range\cite{Wilson:2013prb}, the dashed blue curve calculated from the measured values for $L_{D}$, $M_s$, $K = -0.46$ kJ/m$^3$ and $\mu_0 H_D = 0.85$~T shows that the scattering from a helicoidal state is qualitatively different from the measured asymmetry. The micromagnetic calculations for this field and set of parameters, shown in Fig.~\ref{fig:mumax}(a), indicates that a skyrmion grating is more energetically favourable than the helicoids.  The corresponding neutron spin-asymmetry shown by the solid red curve in Fig.~\ref{fig:PNR}(c) quantitatively reproduces the features in the PNR.  The right-hand panel shows that the difference in the spin-asymmetries for skyrmions and helicoids arises from the difference in the average magnetization in the center of the film.  Since this quantity is a function of the skyrmion density, we have treated $L_{Dy}$ as a fitting parameter in Fig.~\ref{fig:PNR}(c).  A series of skyrmion gratings with varying density were calculated with corresponding spin asymmetries, $\alpha(q_z)$.   A fit to the $\mu_0 H = 0.5$~T data in Fig.~\ref{fig:PNR}(c) gave a value $L_{Dy} = 22$~nm $\pm 7$~nm, as compared to the expected equilibrium value $L_{Dy} = 21$~nm assuming that the DM interaction is isotropic. This estimate is consistent with the more direct measure of the spacing obtained in the SANS experiments described in the next section.

The PNR data in Fig.~\ref{fig:PNR}(d) was collected after the field was reduced to  $\mu_0 H = 0.3$~T -- the region between $H_{\alpha 1}$ and $H_{\beta}$. As the field is lowered below $H_{\alpha 1}$, the peak in the susceptibility suggests a change in the topology.  For this field range in the simulation, the skyrmions are below the strip-out field where they undergo an elliptic instability and elongate into helicoids, as discussed in Ref.~\onlinecite{Bogdanov:1994pss}.  However, the spin asymmetry scattering expected from a 1D helicoidal modulation, represented by the blue dashed line, does not agree with the measured asymmetry.  In a real sample, interfacial twists at the chiral grain boundaries repel skyrmions and restrict their elliptic distortion below the strip-out field. The energy barriers associated the film interfaces and grain boundaries lead to metastable structures consisting of a helicoid-like structures with half-skyrmions at either end. An example of one end of such a structure is shown in Fig.~\ref{fig:mumax}(b). These structures have been observed in Fe$_x$Co$_{1-x}$Si crystals\cite{Yu:2010nat, Milde:2013sci}.  The calculation initially places a single skyrmion straddling across the periodic boundaries, and the relaxation of the magnetic structure causes the skyrmion to elongate along the $y$-direction and fill the entire width of the simulation.  The number of cells in the simulation along the $y$-direction is varied in order to alter the relative amount of skyrmion and helicoid character in the magnetization depth profile.  The red curve in Fig.~\ref{fig:PNR}(d) is for a 20.5~nm wide simulation and corresponds to a single skyrmion.  The best fit to the PNR data is given for a $44\pm17$~nm wide structure, shown by the dotted green curve in Fig.~\ref{fig:PNR}(d).

Finally, Fig.~\ref{fig:PNR}(e) shows that below a field $H_{\beta}$, a pure helicoidal state is recovered.  In the low field regime, the calculated ground state helicoid yields a different spin asymmetry than the data.  The presence of bi-chiral domains creates frustration at the grain boundaries that is responsible for the reported glassy behavior \cite{Karhu:2010prb}. This frustration causes disorder in the phase of the helices in each grain.  The resulting rotation of the net magnetization of the grains away from the applied field produces the reduction in the average moment observed in the data.   We fit the phase, amplitude and wavelength of the helical spin-density wave in Fig.~\ref{fig:PNR}(e).  The right-hand panel shows the fitted magnetic structure. 

\subsection{SANS}
The SANS measurements were performed at the NIST Center for Neutron Research (NCNR) on the NG7 beam line with a neutron wavelength of $0.500 \pm 0.012$~nm.   The sample was mounted in a cryostat with Si windows to minimize the background signal.  The sample-detector distance was set to $4.5$~m to provide a $Q$-range of 0.08 to 0.75 nm$^{-1}$. A circular beam-stop blocked neutrons with scattering vectors $|Q| < 0.063$~nm$^{-1}$.  Eight Si-capped MnSi/Si(111) samples with a $d = 25.2$~nm film thickness were stacked one on top of another with their film normal aligned along the incident neutron beam.

The geometry for the SANS experiment is shown in Fig.~\ref{fig:neutron}(b), where the incident neutrons propagate along the $-z$ direction. The four spin-dependent neutron cross-sections can be expressed as\cite{Maleyev:1963spss,Blume:1963}
\begin{eqnarray}
    I^{\pm \pm} &=&   |F_N(\mathbf{Q}) \pm F_{M,x}(\mathbf{Q})|^2S(\mathbf{Q}),
    \label{eq:I1} \\
    I^{\mp \pm} &=&  |F_{M,z}(\mathbf{Q}) \mp F_{M,y}(\mathbf{Q})|^2S(\mathbf{Q}), 
\label{eq:I2}
\end{eqnarray}
with two superscripts for the intensities to denote the polarization of the out-going and in-coming neutrons relative to the quantization axis set by the field direction (along the $x$-axis). $F_N(\mathbf{Q})$ is the nuclear form factor, $F_{M,i}$ are the components ($i=x,y,z$) of the magnetic form factor for a single skyrmion in the film, and $S(\mathbf{Q})$ is the structure factor.  Only the component of the magnetization perpendicular to the scattering vector $\mathbf{Q}$ contributes to the scattering cross-section: $\mathbf{M}_\perp = \mathbf{M}-(\mathbf{M}\cdot\mathbf{\hat{Q}})\mathbf{\hat{Q}}$ where $\mathbf{\hat{Q}}$ is a unit vector along $\mathbf{Q}$.  Hence the magnetic form factor is a function of this Halpern-Johnson vector, $\mathbf{M}_\perp$,
\begin{equation}
    \mathbf{F_{M}} = \int \mathbf{M}_{\perp}(\mathbf{r}) e^{i\mathbf{Q} \cdot \mathbf{r}} d\mathbf{r}.
\label{eq:FM}
\end{equation}
In the SANS experiment, $Q_z$ is nearly zero and so $\mathbf{F_{M}}$ averages $\mathbf{M}_\perp(\mathbf{r})$ across the film thickness.

\begin{figure}
\includegraphics[width=1.0\columnwidth]{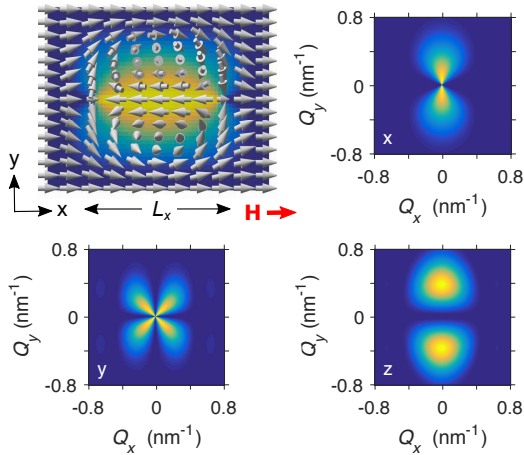}
\caption{
(Color online) An isolated skyrmion taken from a micromagnetic calculation of a disordered arrangement of in-plane skyrmions in a field $H=0.53 H_D$.  The top-left figures shows the spin arrangement in the center of the film.  The colour scale encodes the $x$-component of the magnetization.  The simulation was used to calculate the $x$, $y$, and $z$-components of the Fourier transform of the Halpern-Johnson vector $\mathbf{F}_M(\mathbf{Q})$, shown in the remained 3 panels.
}
\label{fig:HJ}
\end{figure}

\begin{figure}
\includegraphics[width=1.0\columnwidth]{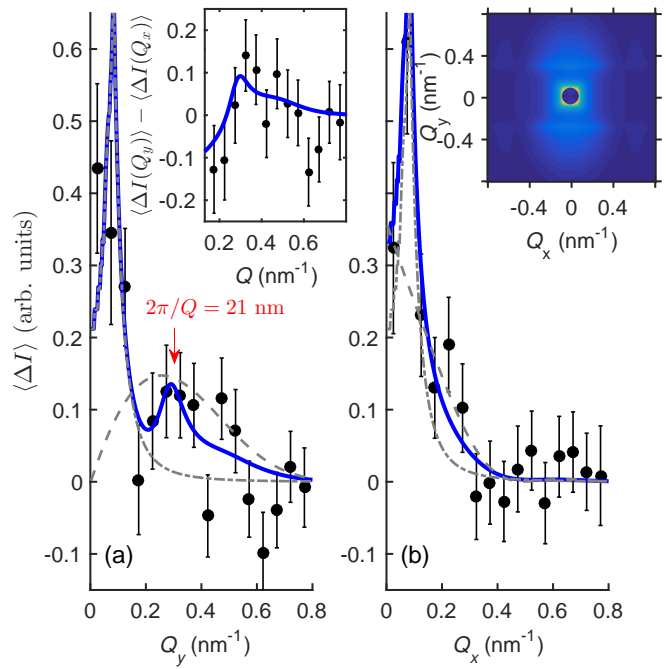}
\caption{
(Color online) SANS measurements of the differential scattering intensity, $\left<\Delta I\right>$, integrated over $Q_x$ in (a) and over $Q_y$ in (b). The error bars are $\pm 1 \sigma$.  The dashed line in (a) is $\left<\Delta I(Q_y)\right>$ for the magnetic form factor obtained from micromagnetic calculations for a single skyrmion shown in Fig.~\ref{fig:HJ}. The dash-dotted line is the contribution from Porod scattering. The blue solid line is the best fit to the data with Eq.~\ref{eq:DI}, with $L_{Dy} = 21$~nm and $\sigma_L = 5.6$~nm.  Inset (a) plots the difference  between $\left<\Delta I\right>$ in Figs. (a) and (b). Inset (b) shows the calculated differential SANS pattern $\Delta I (\mathbf{Q})$ corresponding to the blue curves in (a) and (b).  
}
\label{fig:SANS}
\end{figure}

Due to the small volume in the thin film sample, we used unpolarized neutrons to avoid intensity loss from a polarizer and analyzer.  Therefore to extract the magnetic contrast, we performed a 	differential measurement at $T=30$~K by subtracting a set of SANS intensities measured at $\mu_0 H = 2.0$~T, well above the saturation field, from a set of diffraction patterns measured at $\mu_0 H = 0.45$~T, in the middle of the skyrmion phase.  Data were collected for a total of 21 hours at each field.  The $\mu_0 H = 0.45$~T were recorded after dropping the field from  2.0~T.   The differential intensity, $\Delta I = I(H) - I(H_{sat})$, is proportional to the magnitude of the magnetic form factor,
\begin{equation}
    \Delta I(\mathbf{Q}) =  2 |\mathbf{F_{M}}(\mathbf{Q})|^2 S(\mathbf{Q}),
\label{eq:DI}	
\end{equation}
calculated from Eqs.~(\ref{eq:I1} - \ref{eq:I2}).  We find evidence for an in-plane modulation of the magnetization along the $y$-direction by averaging $\Delta I(\mathbf{Q})$, along $\mathbf{Q}_x$ between $-0.5$ and $+0.5$~nm$^{-1}$, corresponding to $\left<\Delta I(Q_y)\right>$ in Fig.~\ref{fig:SANS}(a):
a weak feature is visible at approximately $Q_y =0.3$~nm$^{-1}$, corresponding to a mean skyrmion spacing $2\pi/0.3$~nm$^{-1} = 21$~nm, as expected from fits to the PNR data.  However, this feature is very broad and indicates a large amount of disorder in the skyrmion array.  As expected, the feature is absent when $\Delta I(\mathbf{Q})$ is averaged along $\mathbf{Q}_y$, as shown by $\left<\Delta I(Q_x)\right>$ in Fig.~\ref{fig:SANS}(b) since the field tends to align the skyrmions along the field.   In addition to the scattering from the skyrmions, there is a relatively large upturn in both $\left<\Delta I(Q_x)\right>$  and $\left<\Delta I(Q_y)\right>$, at low-$Q$.   We argue below that this is due to Porod scattering.  The Porod contribution can be removed by taking the difference between $\left<\Delta I(Q_y)\right>$ and $\left<\Delta I(Q_y)\right>$, leaving the signal from the skyrmions shown in the inset in Fig.~\ref{fig:SANS}(a).

To fit the SANS data, we performed 3D micromagnetic calculations with MuMax3.  The simulation volume was divided into 256 $\times$ 128 $\times$ 64 cubic cells along the $x$-, $y$- and $z$-directions respectively.   We used $\mu_0 H_D = 0.78$~T, $K = -3.8$~kJ/m$^3$ and $M_s = 127$~kA/m in the simulations, estimated from magnetometry measurements, and PNR measurements on similar samples.  To simulate the 25.2~nm thick film we implemented periodic boundary conditions in $x$ and $y$, and free boundaries in $z$.  We created a disordered ensemble of skyrmions by relaxing the spin arrangements from a random arrangement.  Although we cannot discern the detailed nature of the disorder from the experimental data, we choose to fit the data by constructing a simple model from the form factor for a single skyrmion.

Figure~\ref{fig:HJ} shows the spin distribution in the center of the film of one of the skyrmions from the simulation. It shows a conventional skyrmion tube truncated at each end by chiral Block points, similar to the chiral bobbers predicted by Rybakov \cite{Rybakov:2015prl}. These short-length in-plane skyrmions are not stable in isolation, but are frozen in by the disorder.  Small asymmetries in the spin distribution caused by neighbouring skyrmions were removed for clarity, although this makes no difference to the calculation of  $\left<\Delta I(Q_x)\right>$ and  $\left<\Delta I(Q_y)\right>$. The skrymion form factor calculated from the micromagnetic calculation is also shown in Fig.~\ref{fig:HJ}. The symmetry of the z-component  of the skyrmion $M_z(x,y,z) = -M_z(x,-y,z)$ produces a nodal line along $Q_y = 0$ in the $z$-component of $\mathbf{F}_M$. The figure also shows that the other two components of $\mathbf{F}_M$ have nodal lines along $Q_y = 0$ due to the nature of the Halpern-Johnson vector:
\begin{align}
F_{M_x} &= \tilde{M}_x(\mathbf{Q})\hat{Q}_y^2,\\
F_{M_y} &= -\tilde{M}_x(\mathbf{Q})\hat{Q}_x\hat{Q}_y,
\end{align}
where $\tilde{M_x}(\mathbf{Q})$ is the Fourier transform of the $x$-component of the magnetization (the $y$-component,  $\tilde{M_y}(\mathbf{Q})$, integrates to zero across the thickness of the film).  This results in the drop in $\left<\Delta I(Q_y)\right>$ of the form factor at low-$Q$, represented by the dashed grey line in Fig.~\ref{fig:SANS}(a).

The drop in the form factor at $Q=0$ implies that the low-$Q$ upturn is not due to skyrmions and must be due to longer length scale features.  The chiral grain boundaries set a second length scale for the problem that is of the order of several hundred nanometers \cite{Karhu:2011prb}.  Although the differential SANS measurement removes the nuclear contribution from the signal, the grain boundaries imprint their structure on the magnetic texture. The skyrmion lattice phase possesses a net magnetization, as evidenced by PNR, and variations in the direction of the net magnetization of each grain arise in the sample due to the frustration at the grain boundaries\cite{Karhu:2010prb}.  These would be expected to contribute to Porod scattering.  We account for this scattering by including a $Q^{-4}$ term in the model shown by the dash-dotted grey line in Fig.~\ref{fig:SANS}(a) and (b).  Note that the influence of the beam stop is also included in the fit and accounts for the downturn in intensity below $Q=0.063$~nm$^{-1}$.

The disorder in the skrymion lattice was simulated with a structure factor for a 1D paracrystalline material.  Under the assumption that the probability distribution function for the spacing $L_{Dy}$ between neighbouring skyrmions is Gaussian with standard deviation $\sigma_L$, the structure factor takes the following form\cite{Lindenmeyer:1963jap}:
\begin{equation}
S(\mathbf{Q}) = \frac{1 - e^{-Q^2 \sigma_L^2}}
{(1+ e^{-Q^2 \sigma_L^2}) - 2e^{-Q^2 \sigma_L^2/2}\cos(QL_{Dy})}.
\label{eq:SQ}
\end{equation}
The best fit to the integrated differential intensities $\left<\Delta I(Q_y)\right>$ and $\left<\Delta I(Q_x)\right>$ is shown by the blue solid line in Fig.~\ref{fig:SANS} (a) and (b).  The calculated SANS pattern corresponding to the fit is shown in the inset (b). The fit gives a mean skyrmion separation $L_{Dy} = 21$~nm and standard deviation $\sigma_L = 5.6$~nm.   This is in reasonable agreement with the calculated value for the ideal ordered array of skyrmions ($L_{Dy} = 24.7$~nm for $H=0.53 H_D$, and $d = 25.2$~nm).  The differential intensity integrated along $Q_y$, $\left<\Delta I(Q_x)\right>$ is somewhat sensitive to the correlation length of the skyrmions along the direction of the field (see Fig.~\ref{fig:SANS} (b)).  The $\left<\Delta I(Q_x)\right>$ data shows intensity above the Porod scattering term near $Q=0.2$~nm$^{-1}$.  The fit shown in blue corresponds to an average skyrmion length of $L_{x} =14$~nm.  

Although we have chosen the skrymion shown in Fig.~\ref{fig:HJ} in order to be able to fit the skrymion spacing and length, the actual distribution of skyrmions is likely more continuous than implied by the figure since the truncation of the skyrmion tubes introduces additional regions of ferromagnetic order between the skyrmions that increases the overall magnetization of the chiral grain.  Assuming that the disorder in the SANS and PNR samples are the same, our PNR measurements would rule out a spin distribution with a  significant fraction of short skrymions. A collection of meandering skyrmion tubes would also further broaden the modeled peak at $Q_y \approx 0.3$~nm$^{-1}$.   Therefore the value of $L_{x}$ is likely a representation of the correlation length of meandering skyrmions tubes rather than the average length of the tubes.  

We conclude from the SANS data that there is a large degree of disorder in the skyrmion lattice.  Both PNR and SANS give estimates for the mean skyrmion spacing that are within error of one another.   

\section{Discussion}
The PNR and SANS experiments provide direct measures of the magnetic periodicities in our  $25  - 27$~nm thick MnSi films that give further insights into the disputed magnetic phase diagram of MnSi thin films.  In the following, we place this new evidence in the context of previously published results.  We focus the discussion to the portion of the phase diagram corresponding to the decreasing magnetic field branch of the hysteresis loop, along which the PNR and SANS data were collected.

PNR measurements show that these MnSi films are in a ferromagnetic state with surface twists at $\mu_0 H = 0.7$~T, both at $T = 5$~K (Ref.~\onlinecite{Wilson:2013prb}) and $T=25$~K (Fig.~\ref{fig:PNR}(b)) . A comparison of Figs.~\ref{fig:PNR}(b) and (c) demonstrates that PNR can clearly distinguish between a ferromagnetic state and a skyrmionic state.  There are few skyrmions below $T = 10$~K, if any, in the decreasing magnetic field branch: the data indicate for this range of parameters the skyrmions cover approximately $5\% \pm 15\%$ of the film.  The PHE signal for these films at $T = 5$~K (see Fig.~\ref{fig:PD}) does not appear to be due to skyrmions as claimed in Ref.~\onlinecite{Tomoyuki:2015jpsj}, and it is therefore important to consider other possible sources, such as scattering from the frustrated magnetic structures at the chiral domain walls. 

The question of why skyrmions are absent from the low temperature region of the phase diagram despite the predictions of micromagnetic calculations is explained by the kinetics of the problem. As the temperature drops, increases in the magnetocrystalline anisotropy raises the skyrmion nucleation energy.  It is likely that the diminishing thermal fluctuations become too small to nucleate skyrmions, and helicoids remain as metastable objects at intermediate fields.     

There is further disagreement between Refs.~\onlinecite{Wilson:2013prb} and \onlinecite{Tomoyuki:2015jpsj} about the winding of the helicoid at low temperatures.  At $T = 5$~K, PNR provides proves that the system transitions from a twisted ferromagnetic to a set of discrete helicoidal states as the field is lowered.  This  is supported by magnetometry, magnetoresistance and theoretical calculations: Analytical solutions to a finite-size 1D Dzyaloshinskii model explain the thickness dependence of the transition fields and variations in the sign of the low-field magnetoresistance \cite{Wilson:2013prb}. The model shows that for a film thickness $d < L_D$, the system has no peaks in the field dependence of $dM/dH$, and it transitions continuously from a twisted ferromagnet to a partial helix as the $H$ is lowered.  For $L_D \lesssim d \lesssim 2_D$, the system transitions discontinuously from a twisted ferromagnetic to a 1-turn helicoid at a field $H_{h1}$ with a corresponding peak in $dM/dH$.  For $L_D \lesssim d \lesssim 2L_D$, there is an additional transition from a 1-turn state to a 2-turn state at $H_{h2}$. Similarly, for $2L_D \lesssim d \lesssim 3L_D$ the is a third transition from a 2-turn state to a 3-turn state at $H_{h3}$.  

The authors of Ref.~\onlinecite{Tomoyuki:2015jpsj} argue against the presence of discrete helicoids based on the claim that doubling the thickness from 25 nm to 50 nm doesn't increase the number of helicoidal states, as observed by the number of peaks in $dM/dH$.  This, however, is incorrect.  The helicoidal unwinding is inherently non-linear, and the critical fields must be calculated from the Dzyaloshinskii model and compared to the broadening of the transitions due to sample defects, heterogeneity, and grain boundaries.  As the thickness increases, the switching fields all converge upon $H = (\pi^2/16) H_D = 0.617 H_D$ -- the transition field to a ferromagnetic state observed in bulk uniaxial helical magnets.  At a thickness of 3.6 $L_D$, corresponding to the 50 nm thick film in Ref.~\onlinecite{Tomoyuki:2015jpsj}, the 1D model predicts that the magnetic structure transitions from a twisted ferromagnet to the 3.5 turn helical ground state via the nucleation of helicoid turns at fields $H_{h1} = 0.617 H_D$, $H_{h2} = 0.597 H_D$ and $H_{h3} = 0.439 H_D$. But in practice $H_{h1}$ and $H_{h2}$ cannot be distinguished in a 50 nm film because the width of the peaks in $dM/dH$ (of the order of 0.08~T in Fig.~\ref{fig:hys}) is much larger than $H_{h1}-H_{h2}$.  We mention in passing that discrete states have also been observed in micron sized CrNb$_3$S$_6$ crytals \cite{Togawa:2015prb}, but are reported to be absent from FeGe/Si(111) for reasons that are not clear \cite{Kanazawa:2016prb}.

As the temperature is raised, the MnSi films depart from the simple discrete helicoid picture. The change in the $dM/dH$ data with the appearance of an additional peak heralds a significant change to the magnetic structure for $T \gtrsim 12$~K \cite{Wilson:2012prb}. The pair of large peaks at $H_{\alpha1}$ and $H_{\alpha2}$ in $dM/dH$ observed in Fig.~\ref{fig:hys} demarcate the boundaries of the new magnetic phase shown in Fig.~\ref{fig:PD}, which is the main subject of this paper.   The same $dM/dH$ features are observed by Tomoyuki \emph{et al.} (see Fig. 4(e) of Ref.~\onlinecite{Tomoyuki:2015jpsj}), although the authors were dismissive of the ``tiny anomalies in $M$".  We point out that while the changes in $M$ are subtle, the peaks in the susceptibilities at $H_{\alpha1}$ and $H_{\alpha2}$ are not:  The peaks in $dM/dH$ in Fig.~\ref{fig:hys} are about a factor of 2 times higher than $dM/dH$ in the neighboring phases, and very similar in magnitude to bulk MnSi in the middle of the A-phase (compare Fig.~\ref{fig:hys} with Fig.3(c) in Ref.~\onlinecite{Bauer:2012prb}).  

In this paper we have used neutrons to probe the magnetic structure of the phase shown in red in Fig.~\ref{fig:PD} in all three dimensions. The PNR measurements presented in Fig.~\ref{fig:PNR}(c) show that the magnetic texture continues to be highly ordered across the depth of the film, and that there are solitons localized in the center of the film that are distinct from the 1D helicoidal modulations at low temperatures.  The numerical calculations of the skyrmion phase give a good match to the thickness dependence of $M$ and by fitting the simulations to Fig.~\ref{fig:PNR}(c), we find that the reduction in the average magnetization in the center of the film corresponds to an in-plane skyrmion spacing of $L_{Dy} = 22$~nm $\pm 7$~nm.   The complementary SANS results confirm that the solitons in the middle of the film are multi-dimensional: the skyrmions have an average spacing of $21$~nm $\pm 6$~nm, in agreement with the estimate from PNR.  While the in-plane value for $L_{Dy}$ is consistent with numerical calculations of a skyrmion phase, the diffraction shows that it is highly-disordered in the plane.  The combination of frustration at the irregularly shaped chiral domain boundaries and thermal fluctuations are likely factors that contribute to the destruction of long range order in-plane.  Repulsion between the skyrmions and the surface twists provide a confining potential \cite{Nagaosa:2013nn, Meynell:2014prb1} that drives them to the middle of the film and leads to the creation of a 2D skyrmion glass.  It is not clear why there is no PHE in this region of the phase diagram.   However, the effect, which depends on both band structure and anisotropic electron scattering, is not fully understood.  In the case of MnSi there are large variations in the size and sign of the PHE depending on the direction of the magnetic field and the skrymion signal nearly vanishes for $\mathbf{H} \| [111]$ \cite{Tomoyuki:2015jpsj}.  A more in-depth understanding of the electron scattering in MnSi in all the magnetic textures is needed to properly interpret the electron-transport measurements.
\\

\section{Conclusion}
Neutron scattering experiments presented in this paper provide valuable information about the magnetic structure of MnSi thin films to resolve two opposing interpretations of previous transport and magnetometry measurements.  The neutron experiments confirm the interpretation of the MnSi phase diagram shown in Fig.~\ref{fig:PD} and prove the existence of multidimensional solitons in the center of the MnSi films for in-plane magnetic fields.  The SANS data reveal that the skyrmion phase possesses a large degree of disorder, possibly due to the chiral grain boundaries.  This work, together with previous transport measurements \cite{Meynell:2014prb2, Tomoyuki:2015jpsj} will hopefully motivate further theoretical work to understand the Hall effect and the planar Hall effect in B20 thin films.

\section{Acknowledgments}
We would like to thank Alex Bogdanov and Andrey Leonov for many helpful discussions.  We would also like to thank Filipp Rybakov for providing preliminary calculations. T.L.M., M.N.W. and S.A.M. acknowledge support from NSERC. This research was enabled in part by support provided by WestGrid (www.westgrid.ca) and Compute Canada (www.computecanada.ca)

%


\begin{thebibliography}{53}%
\makeatletter
\providecommand \@ifxundefined [1]{%
 \@ifx{#1\undefined}
}%
\providecommand \@ifnum [1]{%
 \ifnum #1\expandafter \@firstoftwo
 \else \expandafter \@secondoftwo
 \fi
}%
\providecommand \@ifx [1]{%
 \ifx #1\expandafter \@firstoftwo
 \else \expandafter \@secondoftwo
 \fi
}%
\providecommand \natexlab [1]{#1}%
\providecommand \enquote  [1]{``#1''}%
\providecommand \bibnamefont  [1]{#1}%
\providecommand \bibfnamefont [1]{#1}%
\providecommand \citenamefont [1]{#1}%
\providecommand \href@noop [0]{\@secondoftwo}%
\providecommand \href [0]{\begingroup \@sanitize@url \@href}%
\providecommand \@href[1]{\@@startlink{#1}\@@href}%
\providecommand \@@href[1]{\endgroup#1\@@endlink}%
\providecommand \@sanitize@url [0]{\catcode `\\12\catcode `\$12\catcode
  `\&12\catcode `\#12\catcode `\^12\catcode `\_12\catcode `\%12\relax}%
\providecommand \@@startlink[1]{}%
\providecommand \@@endlink[0]{}%
\providecommand \url  [0]{\begingroup\@sanitize@url \@url }%
\providecommand \@url [1]{\endgroup\@href {#1}{\urlprefix }}%
\providecommand \urlprefix  [0]{URL }%
\providecommand \Eprint [0]{\href }%
\providecommand \doibase [0]{http://dx.doi.org/}%
\providecommand \selectlanguage [0]{\@gobble}%
\providecommand \bibinfo  [0]{\@secondoftwo}%
\providecommand \bibfield  [0]{\@secondoftwo}%
\providecommand \translation [1]{[#1]}%
\providecommand \BibitemOpen [0]{}%
\providecommand \bibitemStop [0]{}%
\providecommand \bibitemNoStop [0]{.\EOS\space}%
\providecommand \EOS [0]{\spacefactor3000\relax}%
\providecommand \BibitemShut  [1]{\csname bibitem#1\endcsname}%
\let\auto@bib@innerbib\@empty
\bibitem [{\citenamefont {Dzyaloshinsky}(1958)}]{Dzyaloshinskii:1958jpcs}%
  \BibitemOpen
  \bibfield  {author} {\bibinfo {author} {\bibfnamefont {I.}~\bibnamefont
  {Dzyaloshinsky}},\ }\href {\doibase
  http://dx.doi.org/10.1016/0022-3697(58)90076-3} {\bibfield  {journal}
  {\bibinfo  {journal} {J. Phys. Chem. Solids}\ }\textbf {\bibinfo {volume}
  {4}},\ \bibinfo {pages} {241 } (\bibinfo {year} {1958})}\BibitemShut
  {NoStop}%
\bibitem [{\citenamefont {Moriya}(1960)}]{Moriya:1960pr}%
  \BibitemOpen
  \bibfield  {author} {\bibinfo {author} {\bibfnamefont {T.}~\bibnamefont
  {Moriya}},\ }\href {\doibase 10.1103/PhysRev.120.91} {\bibfield  {journal}
  {\bibinfo  {journal} {Phys. Rev.}\ }\textbf {\bibinfo {volume} {120}},\
  \bibinfo {pages} {91} (\bibinfo {year} {1960})}\BibitemShut {NoStop}%
\bibitem [{\citenamefont {Fert}(1990)}]{Fert:1990msf}%
  \BibitemOpen
  \bibfield  {author} {\bibinfo {author} {\bibfnamefont {A.}~\bibnamefont
  {Fert}},\ }\href@noop {} {\bibfield  {journal} {\bibinfo  {journal} {Mater.
  Sci. Forum}\ }\textbf {\bibinfo {volume} {59-60}},\ \bibinfo {pages} {439}
  (\bibinfo {year} {1990})}\BibitemShut {NoStop}%
\bibitem [{\citenamefont {Dzyaloshinskii}(1964)}]{Dzyaloshinskii:1964zetf-2}%
  \BibitemOpen
  \bibfield  {author} {\bibinfo {author} {\bibfnamefont {I.~E.}\ \bibnamefont
  {Dzyaloshinskii}},\ }\href@noop {} {\bibfield  {journal} {\bibinfo  {journal}
  {Zh. \'Eksp. Teor. Fiz.}\ }\textbf {\bibinfo {volume} {47}},\ \bibinfo
  {pages} {336 } (\bibinfo {year} {1964})},\ \bibinfo {note} {[ Sov. Phys. JETP
  \textbf{20}, 223 (1965) ]}\BibitemShut {NoStop}%
\bibitem [{\citenamefont {Bogdanov}\ and\ \citenamefont
  {Yablonskii}(1989)}]{Bogdanov:1989jetp}%
  \BibitemOpen
  \bibfield  {author} {\bibinfo {author} {\bibfnamefont {A.~N.}\ \bibnamefont
  {Bogdanov}}\ and\ \bibinfo {author} {\bibfnamefont {D.~A.}\ \bibnamefont
  {Yablonskii}},\ }\href
  {http://www.jetp.ac.ru/cgi-bin/e/index/e/68/1/p101?a=list} {\bibfield
  {journal} {\bibinfo  {journal} {Zh. \'Eksp. Teor. Fiz.}\ }\textbf {\bibinfo
  {volume} {95}},\ \bibinfo {pages} {178} (\bibinfo {year} {1989})},\ \bibinfo
  {note} {[ Sov. Phys. JETP \textbf{68}, 101 (1989) ]}\BibitemShut {NoStop}%
\bibitem [{\citenamefont {Bogdanov}\ and\ \citenamefont
  {Hubert}(1994{\natexlab{a}})}]{Bogdanov:1994jmmm}%
  \BibitemOpen
  \bibfield  {author} {\bibinfo {author} {\bibfnamefont {A.}~\bibnamefont
  {Bogdanov}}\ and\ \bibinfo {author} {\bibfnamefont {A.}~\bibnamefont
  {Hubert}},\ }\href {\doibase DOI: 10.1016/0304-8853(94)90046-9} {\bibfield
  {journal} {\bibinfo  {journal} {J. Magn. Magn. Mater.}\ }\textbf {\bibinfo
  {volume} {138}},\ \bibinfo {pages} {255 } (\bibinfo {year}
  {1994}{\natexlab{a}})}\BibitemShut {NoStop}%
\bibitem [{\citenamefont {Kiselev}\ \emph {et~al.}(2011)\citenamefont
  {Kiselev}, \citenamefont {Bogdanov}, \citenamefont {Sch\"{a}fer},\ and\
  \citenamefont {R\"{o}\ss{}ler}}]{Kiselev:2011jpd}%
  \BibitemOpen
  \bibfield  {author} {\bibinfo {author} {\bibfnamefont {N.~S.}\ \bibnamefont
  {Kiselev}}, \bibinfo {author} {\bibfnamefont {A.~N.}\ \bibnamefont
  {Bogdanov}}, \bibinfo {author} {\bibfnamefont {R.}~\bibnamefont
  {Sch\"{a}fer}}, \ and\ \bibinfo {author} {\bibfnamefont {U.~K.}\ \bibnamefont
  {R\"{o}\ss{}ler}},\ }\href {http://stacks.iop.org/0022-3727/44/i=39/a=392001}
  {\bibfield  {journal} {\bibinfo  {journal} {J. Phys. D: Appl. Phys.}\
  }\textbf {\bibinfo {volume} {44}},\ \bibinfo {pages} {392001} (\bibinfo
  {year} {2011})}\BibitemShut {NoStop}%
\bibitem [{\citenamefont {Sampaio}\ \emph {et~al.}(2013)\citenamefont
  {Sampaio}, \citenamefont {Cros}, \citenamefont {Rohart}, \citenamefont
  {Thiaville},\ and\ \citenamefont {Fert}}]{Sampaio:2013nn}%
  \BibitemOpen
  \bibfield  {author} {\bibinfo {author} {\bibfnamefont {J.}~\bibnamefont
  {Sampaio}}, \bibinfo {author} {\bibfnamefont {V.}~\bibnamefont {Cros}},
  \bibinfo {author} {\bibfnamefont {S.}~\bibnamefont {Rohart}}, \bibinfo
  {author} {\bibfnamefont {A.}~\bibnamefont {Thiaville}}, \ and\ \bibinfo
  {author} {\bibfnamefont {A.}~\bibnamefont {Fert}},\ }\href
  {http://dx.doi.org/10.1038/nnano.2013.210} {\bibfield  {journal} {\bibinfo
  {journal} {Nat Nano}\ }\textbf {\bibinfo {volume} {8}},\ \bibinfo {pages}
  {839} (\bibinfo {year} {2013})}\BibitemShut {NoStop}%
\bibitem [{\citenamefont {Nagaosa}\ and\ \citenamefont
  {Tokura}(2013)}]{Nagaosa:2013nn}%
  \BibitemOpen
  \bibfield  {author} {\bibinfo {author} {\bibfnamefont {N.}~\bibnamefont
  {Nagaosa}}\ and\ \bibinfo {author} {\bibfnamefont {Y.}~\bibnamefont
  {Tokura}},\ }\href {http://dx.doi.org/10.1038/nnano.2013.243} {\bibfield
  {journal} {\bibinfo  {journal} {Nat Nano}\ }\textbf {\bibinfo {volume} {8}},\
  \bibinfo {pages} {899} (\bibinfo {year} {2013})}\BibitemShut {NoStop}%
\bibitem [{\citenamefont {Bogdanov}(2000)}]{Bogdanov:2000jetp}%
  \BibitemOpen
  \bibfield  {author} {\bibinfo {author} {\bibfnamefont {A.~N.}\ \bibnamefont
  {Bogdanov}},\ }\href {http://dx.doi.org/10.1134/1.568286} {\bibfield
  {journal} {\bibinfo  {journal} {Journal of Experimental and Theoretical
  Physics Letters}\ }\textbf {\bibinfo {volume} {71}},\ \bibinfo {pages} {85}
  (\bibinfo {year} {2000})}\BibitemShut {NoStop}%
\bibitem [{\citenamefont {Leonov}\ \emph {et~al.}(2014)\citenamefont {Leonov},
  \citenamefont {Dragunov}, \citenamefont {R\"o\ss{}ler},\ and\ \citenamefont
  {Bogdanov}}]{Leonov:2014pre}%
  \BibitemOpen
  \bibfield  {author} {\bibinfo {author} {\bibfnamefont {A.~O.}\ \bibnamefont
  {Leonov}}, \bibinfo {author} {\bibfnamefont {I.~E.}\ \bibnamefont
  {Dragunov}}, \bibinfo {author} {\bibfnamefont {U.~K.}\ \bibnamefont
  {R\"o\ss{}ler}}, \ and\ \bibinfo {author} {\bibfnamefont {A.~N.}\
  \bibnamefont {Bogdanov}},\ }\href {\doibase 10.1103/PhysRevE.90.042502}
  {\bibfield  {journal} {\bibinfo  {journal} {Phys. Rev. E}\ }\textbf {\bibinfo
  {volume} {90}},\ \bibinfo {pages} {042502} (\bibinfo {year}
  {2014})}\BibitemShut {NoStop}%
\bibitem [{\citenamefont {Ackerman}\ \emph {et~al.}(2014)\citenamefont
  {Ackerman}, \citenamefont {Trivedi}, \citenamefont {Senyuk}, \citenamefont
  {van~de Lagemaat},\ and\ \citenamefont {Smalyukh}}]{Ackerman:2014pre}%
  \BibitemOpen
  \bibfield  {author} {\bibinfo {author} {\bibfnamefont {P.~J.}\ \bibnamefont
  {Ackerman}}, \bibinfo {author} {\bibfnamefont {R.~P.}\ \bibnamefont
  {Trivedi}}, \bibinfo {author} {\bibfnamefont {B.}~\bibnamefont {Senyuk}},
  \bibinfo {author} {\bibfnamefont {J.}~\bibnamefont {van~de Lagemaat}}, \ and\
  \bibinfo {author} {\bibfnamefont {I.~I.}\ \bibnamefont {Smalyukh}},\ }\href
  {\doibase 10.1103/PhysRevE.90.012505} {\bibfield  {journal} {\bibinfo
  {journal} {Phys. Rev. E}\ }\textbf {\bibinfo {volume} {90}},\ \bibinfo
  {pages} {012505} (\bibinfo {year} {2014})}\BibitemShut {NoStop}%
\bibitem [{\citenamefont {R\"{o}\ss{}ler}\ \emph {et~al.}(2006)\citenamefont
  {R\"{o}\ss{}ler}, \citenamefont {Bogdanov},\ and\ \citenamefont
  {Pfleiderer}}]{Rossler:2006nat}%
  \BibitemOpen
  \bibfield  {author} {\bibinfo {author} {\bibfnamefont {U.~K.}\ \bibnamefont
  {R\"{o}\ss{}ler}}, \bibinfo {author} {\bibfnamefont {A.~N.}\ \bibnamefont
  {Bogdanov}}, \ and\ \bibinfo {author} {\bibfnamefont {C.}~\bibnamefont
  {Pfleiderer}},\ }\href {\doibase AUG 17} {\bibfield  {journal} {\bibinfo
  {journal} {Nature}\ }\textbf {\bibinfo {volume} {442}},\ \bibinfo {pages}
  {797} (\bibinfo {year} {2006})}\BibitemShut {NoStop}%
\bibitem [{\citenamefont {M\"{u}hlbauer}\ \emph {et~al.}(2009)\citenamefont
  {M\"{u}hlbauer}, \citenamefont {Binz}, \citenamefont {Jonietz}, \citenamefont
  {Pfleiderer}, \citenamefont {Rosch}, \citenamefont {Neubauer}, \citenamefont
  {Georgii},\ and\ \citenamefont {B\"{o}ni}}]{Muhlbauer:2009sci}%
  \BibitemOpen
  \bibfield  {author} {\bibinfo {author} {\bibfnamefont {S.}~\bibnamefont
  {M\"{u}hlbauer}}, \bibinfo {author} {\bibfnamefont {B.}~\bibnamefont {Binz}},
  \bibinfo {author} {\bibfnamefont {F.}~\bibnamefont {Jonietz}}, \bibinfo
  {author} {\bibfnamefont {C.}~\bibnamefont {Pfleiderer}}, \bibinfo {author}
  {\bibfnamefont {A.}~\bibnamefont {Rosch}}, \bibinfo {author} {\bibfnamefont
  {A.}~\bibnamefont {Neubauer}}, \bibinfo {author} {\bibfnamefont
  {R.}~\bibnamefont {Georgii}}, \ and\ \bibinfo {author} {\bibfnamefont
  {P.}~\bibnamefont {B\"{o}ni}},\ }\href {\doibase FEB 13} {\bibfield
  {journal} {\bibinfo  {journal} {Science}\ }\textbf {\bibinfo {volume}
  {323}},\ \bibinfo {pages} {915} (\bibinfo {year} {2009})}\BibitemShut
  {NoStop}%
\bibitem [{\citenamefont {Pappas}\ \emph {et~al.}(2009)\citenamefont {Pappas},
  \citenamefont {Leli\`{e}vre-Berna}, \citenamefont {Falus}, \citenamefont
  {Bentley}, \citenamefont {Moskvin}, \citenamefont {Grigoriev}, \citenamefont
  {Fouquet},\ and\ \citenamefont {Farago}}]{Pappas:2009prl}%
  \BibitemOpen
  \bibfield  {author} {\bibinfo {author} {\bibfnamefont {C.}~\bibnamefont
  {Pappas}}, \bibinfo {author} {\bibfnamefont {E.}~\bibnamefont
  {Leli\`{e}vre-Berna}}, \bibinfo {author} {\bibfnamefont {P.}~\bibnamefont
  {Falus}}, \bibinfo {author} {\bibfnamefont {P.~M.}\ \bibnamefont {Bentley}},
  \bibinfo {author} {\bibfnamefont {E.}~\bibnamefont {Moskvin}}, \bibinfo
  {author} {\bibfnamefont {S.}~\bibnamefont {Grigoriev}}, \bibinfo {author}
  {\bibfnamefont {P.}~\bibnamefont {Fouquet}}, \ and\ \bibinfo {author}
  {\bibfnamefont {B.}~\bibnamefont {Farago}},\ }\href
  {http://link.aps.org/doi/10.1103/PhysRevLett.102.197202} {\bibfield
  {journal} {\bibinfo  {journal} {Phys. Rev. Lett.}\ }\textbf {\bibinfo
  {volume} {102}},\ \bibinfo {pages} {197202} (\bibinfo {year}
  {2009})}\BibitemShut {NoStop}%
\bibitem [{\citenamefont {Wilhelm}\ \emph {et~al.}(2011)\citenamefont
  {Wilhelm}, \citenamefont {Baenitz}, \citenamefont {Schmidt}, \citenamefont
  {R\"o\ss{}ler}, \citenamefont {Leonov},\ and\ \citenamefont
  {Bogdanov}}]{Wilhelm:2011prl}%
  \BibitemOpen
  \bibfield  {author} {\bibinfo {author} {\bibfnamefont {H.}~\bibnamefont
  {Wilhelm}}, \bibinfo {author} {\bibfnamefont {M.}~\bibnamefont {Baenitz}},
  \bibinfo {author} {\bibfnamefont {M.}~\bibnamefont {Schmidt}}, \bibinfo
  {author} {\bibfnamefont {U.~K.}\ \bibnamefont {R\"o\ss{}ler}}, \bibinfo
  {author} {\bibfnamefont {A.~A.}\ \bibnamefont {Leonov}}, \ and\ \bibinfo
  {author} {\bibfnamefont {A.~N.}\ \bibnamefont {Bogdanov}},\ }\href {\doibase
  10.1103/PhysRevLett.107.127203} {\bibfield  {journal} {\bibinfo  {journal}
  {Phys. Rev. Lett.}\ }\textbf {\bibinfo {volume} {107}},\ \bibinfo {pages}
  {127203} (\bibinfo {year} {2011})}\BibitemShut {NoStop}%
\bibitem [{\citenamefont {Butenko}\ \emph {et~al.}(2010)\citenamefont
  {Butenko}, \citenamefont {Leonov}, \citenamefont {R\"{o}\ss{}ler},\ and\
  \citenamefont {Bogdanov}}]{Butenko:2010prb}%
  \BibitemOpen
  \bibfield  {author} {\bibinfo {author} {\bibfnamefont {A.~B.}\ \bibnamefont
  {Butenko}}, \bibinfo {author} {\bibfnamefont {A.~A.}\ \bibnamefont {Leonov}},
  \bibinfo {author} {\bibfnamefont {U.~K.}\ \bibnamefont {R\"{o}\ss{}ler}}, \
  and\ \bibinfo {author} {\bibfnamefont {A.~N.}\ \bibnamefont {Bogdanov}},\
  }\href {\doibase 10.1103/PhysRevB.82.052403} {\bibfield  {journal} {\bibinfo
  {journal} {Phys. Rev. B}\ }\textbf {\bibinfo {volume} {82}},\ \bibinfo
  {pages} {052403} (\bibinfo {year} {2010})}\BibitemShut {NoStop}%
\bibitem [{\citenamefont {Vousden}\ \emph {et~al.}(2016)\citenamefont
  {Vousden}, \citenamefont {Albert}, \citenamefont {Beg}, \citenamefont
  {Bisotti}, \citenamefont {Carey}, \citenamefont {Chernyshenko}, \citenamefont
  {Cort\'{e}s-Ortuno}, \citenamefont {Wang}, \citenamefont {Hovorka},
  \citenamefont {Marrows},\ and\ \citenamefont {Fangohr}}]{Vousden:2016apl}%
  \BibitemOpen
  \bibfield  {author} {\bibinfo {author} {\bibfnamefont {M.}~\bibnamefont
  {Vousden}}, \bibinfo {author} {\bibfnamefont {M.}~\bibnamefont {Albert}},
  \bibinfo {author} {\bibfnamefont {M.}~\bibnamefont {Beg}}, \bibinfo {author}
  {\bibfnamefont {M.-A.}\ \bibnamefont {Bisotti}}, \bibinfo {author}
  {\bibfnamefont {R.}~\bibnamefont {Carey}}, \bibinfo {author} {\bibfnamefont
  {D.}~\bibnamefont {Chernyshenko}}, \bibinfo {author} {\bibfnamefont
  {D.}~\bibnamefont {Cort\'{e}s-Ortuno}}, \bibinfo {author} {\bibfnamefont
  {W.}~\bibnamefont {Wang}}, \bibinfo {author} {\bibfnamefont {O.}~\bibnamefont
  {Hovorka}}, \bibinfo {author} {\bibfnamefont {C.~H.}\ \bibnamefont
  {Marrows}}, \ and\ \bibinfo {author} {\bibfnamefont {H.}~\bibnamefont
  {Fangohr}},\ }\href {\doibase http://dx.doi.org/10.1063/1.4945262} {\bibfield
   {journal} {\bibinfo  {journal} {Appl. Phys. Lett.}\ }\textbf {\bibinfo
  {volume} {108}},\ \bibinfo {pages} {132406} (\bibinfo {year}
  {2016})}\BibitemShut {NoStop}%
\bibitem [{\citenamefont {Rybakov}\ \emph {et~al.}(2013)\citenamefont
  {Rybakov}, \citenamefont {Borisov},\ and\ \citenamefont
  {Bogdanov}}]{Rybakov:2013prb}%
  \BibitemOpen
  \bibfield  {author} {\bibinfo {author} {\bibfnamefont {F.~N.}\ \bibnamefont
  {Rybakov}}, \bibinfo {author} {\bibfnamefont {A.~B.}\ \bibnamefont
  {Borisov}}, \ and\ \bibinfo {author} {\bibfnamefont {A.~N.}\ \bibnamefont
  {Bogdanov}},\ }\href {\doibase 10.1103/PhysRevB.87.094424} {\bibfield
  {journal} {\bibinfo  {journal} {Phys. Rev. B}\ }\textbf {\bibinfo {volume}
  {87}},\ \bibinfo {pages} {094424} (\bibinfo {year} {2013})}\BibitemShut
  {NoStop}%
\bibitem [{\citenamefont {Rybakov}\ \emph {et~al.}(2015)\citenamefont
  {Rybakov}, \citenamefont {Borisov}, \citenamefont {Bl\"ugel},\ and\
  \citenamefont {Kiselev}}]{Rybakov:2015prl}%
  \BibitemOpen
  \bibfield  {author} {\bibinfo {author} {\bibfnamefont {F.~N.}\ \bibnamefont
  {Rybakov}}, \bibinfo {author} {\bibfnamefont {A.~B.}\ \bibnamefont
  {Borisov}}, \bibinfo {author} {\bibfnamefont {S.}~\bibnamefont {Bl\"ugel}}, \
  and\ \bibinfo {author} {\bibfnamefont {N.~S.}\ \bibnamefont {Kiselev}},\
  }\href {\doibase 10.1103/PhysRevLett.115.117201} {\bibfield  {journal}
  {\bibinfo  {journal} {Phys. Rev. Lett.}\ }\textbf {\bibinfo {volume} {115}},\
  \bibinfo {pages} {117201} (\bibinfo {year} {2015})}\BibitemShut {NoStop}%
\bibitem [{\citenamefont {Yu}\ \emph {et~al.}(2010)\citenamefont {Yu},
  \citenamefont {Onose}, \citenamefont {Kanazawa}, \citenamefont {Park},
  \citenamefont {Han}, \citenamefont {Matsui}, \citenamefont {Nagaosa},\ and\
  \citenamefont {Tokura}}]{Yu:2010nat}%
  \BibitemOpen
  \bibfield  {author} {\bibinfo {author} {\bibfnamefont {X.~Z.}\ \bibnamefont
  {Yu}}, \bibinfo {author} {\bibfnamefont {Y.}~\bibnamefont {Onose}}, \bibinfo
  {author} {\bibfnamefont {N.}~\bibnamefont {Kanazawa}}, \bibinfo {author}
  {\bibfnamefont {J.~H.}\ \bibnamefont {Park}}, \bibinfo {author}
  {\bibfnamefont {J.~H.}\ \bibnamefont {Han}}, \bibinfo {author} {\bibfnamefont
  {Y.}~\bibnamefont {Matsui}}, \bibinfo {author} {\bibfnamefont
  {N.}~\bibnamefont {Nagaosa}}, \ and\ \bibinfo {author} {\bibfnamefont
  {Y.}~\bibnamefont {Tokura}},\ }\href {\doibase 10.1038/nature09124}
  {\bibfield  {journal} {\bibinfo  {journal} {Nature}\ }\textbf {\bibinfo
  {volume} {465}},\ \bibinfo {pages} {901} (\bibinfo {year}
  {2010})}\BibitemShut {NoStop}%
\bibitem [{\citenamefont {Yu}\ \emph {et~al.}(2011)\citenamefont {Yu},
  \citenamefont {Kanazawa}, \citenamefont {Onose}, \citenamefont {Kimoto},
  \citenamefont {Zhang}, \citenamefont {Ishiwata}, \citenamefont {Matsui},\
  and\ \citenamefont {Tokura}}]{Yu:2011nm}%
  \BibitemOpen
  \bibfield  {author} {\bibinfo {author} {\bibfnamefont {X.~Z.}\ \bibnamefont
  {Yu}}, \bibinfo {author} {\bibfnamefont {N.}~\bibnamefont {Kanazawa}},
  \bibinfo {author} {\bibfnamefont {Y.}~\bibnamefont {Onose}}, \bibinfo
  {author} {\bibfnamefont {K.}~\bibnamefont {Kimoto}}, \bibinfo {author}
  {\bibfnamefont {W.~Z.}\ \bibnamefont {Zhang}}, \bibinfo {author}
  {\bibfnamefont {S.}~\bibnamefont {Ishiwata}}, \bibinfo {author}
  {\bibfnamefont {Y.}~\bibnamefont {Matsui}}, \ and\ \bibinfo {author}
  {\bibfnamefont {Y.}~\bibnamefont {Tokura}},\ }\href {\doibase
  10.1038/nmat2916} {\bibfield  {journal} {\bibinfo  {journal} {Nat. Mater.}\
  }\textbf {\bibinfo {volume} {10}},\ \bibinfo {pages} {106} (\bibinfo {year}
  {2011})}\BibitemShut {NoStop}%
\bibitem [{\citenamefont {Tonomura}\ \emph {et~al.}(2012)\citenamefont
  {Tonomura}, \citenamefont {Yu}, \citenamefont {Yanagisawa}, \citenamefont
  {Matsuda}, \citenamefont {Onose}, \citenamefont {Kanazawa}, \citenamefont
  {Park},\ and\ \citenamefont {Tokura}}]{Tonomura:2012nl}%
  \BibitemOpen
  \bibfield  {author} {\bibinfo {author} {\bibfnamefont {A.}~\bibnamefont
  {Tonomura}}, \bibinfo {author} {\bibfnamefont {X.}~\bibnamefont {Yu}},
  \bibinfo {author} {\bibfnamefont {K.}~\bibnamefont {Yanagisawa}}, \bibinfo
  {author} {\bibfnamefont {T.}~\bibnamefont {Matsuda}}, \bibinfo {author}
  {\bibfnamefont {Y.}~\bibnamefont {Onose}}, \bibinfo {author} {\bibfnamefont
  {N.}~\bibnamefont {Kanazawa}}, \bibinfo {author} {\bibfnamefont {H.~S.}\
  \bibnamefont {Park}}, \ and\ \bibinfo {author} {\bibfnamefont
  {Y.}~\bibnamefont {Tokura}},\ }\href {http://dx.doi.org/10.1021/nl300073m}
  {\bibfield  {journal} {\bibinfo  {journal} {Nano Lett.}\ }\textbf {\bibinfo
  {volume} {12}},\ \bibinfo {pages} {1673} (\bibinfo {year}
  {2012})}\BibitemShut {NoStop}%
\bibitem [{\citenamefont {Du}\ \emph {et~al.}(2015)\citenamefont {Du},
  \citenamefont {Che}, \citenamefont {Kong}, \citenamefont {Zhao},
  \citenamefont {Jin}, \citenamefont {Wang}, \citenamefont {Yang},
  \citenamefont {Ning}, \citenamefont {Li}, \citenamefont {Jin}, \citenamefont
  {Chen}, \citenamefont {Zang}, \citenamefont {Zhang},\ and\ \citenamefont
  {Tian}}]{Du:2015nc2}%
  \BibitemOpen
  \bibfield  {author} {\bibinfo {author} {\bibfnamefont {H.}~\bibnamefont
  {Du}}, \bibinfo {author} {\bibfnamefont {R.}~\bibnamefont {Che}}, \bibinfo
  {author} {\bibfnamefont {L.}~\bibnamefont {Kong}}, \bibinfo {author}
  {\bibfnamefont {X.}~\bibnamefont {Zhao}}, \bibinfo {author} {\bibfnamefont
  {C.}~\bibnamefont {Jin}}, \bibinfo {author} {\bibfnamefont {C.}~\bibnamefont
  {Wang}}, \bibinfo {author} {\bibfnamefont {J.}~\bibnamefont {Yang}}, \bibinfo
  {author} {\bibfnamefont {W.}~\bibnamefont {Ning}}, \bibinfo {author}
  {\bibfnamefont {R.}~\bibnamefont {Li}}, \bibinfo {author} {\bibfnamefont
  {C.}~\bibnamefont {Jin}}, \bibinfo {author} {\bibfnamefont {X.}~\bibnamefont
  {Chen}}, \bibinfo {author} {\bibfnamefont {J.}~\bibnamefont {Zang}}, \bibinfo
  {author} {\bibfnamefont {Y.}~\bibnamefont {Zhang}}, \ and\ \bibinfo {author}
  {\bibfnamefont {M.}~\bibnamefont {Tian}},\ }\href
  {http://dx.doi.org/10.1038/ncomms9504} {\bibfield  {journal} {\bibinfo
  {journal} {Nat Commun}\ }\textbf {\bibinfo {volume} {6}} (\bibinfo {year}
  {2015})}\BibitemShut {NoStop}%
\bibitem [{\citenamefont {Yu}\ \emph {et~al.}(2015)\citenamefont {Yu},
  \citenamefont {Kikkawa}, \citenamefont {Morikawa}, \citenamefont {Shibata},
  \citenamefont {Tokunaga}, \citenamefont {Taguchi},\ and\ \citenamefont
  {Tokura}}]{Yu:2015prb}%
  \BibitemOpen
  \bibfield  {author} {\bibinfo {author} {\bibfnamefont {X.}~\bibnamefont
  {Yu}}, \bibinfo {author} {\bibfnamefont {A.}~\bibnamefont {Kikkawa}},
  \bibinfo {author} {\bibfnamefont {D.}~\bibnamefont {Morikawa}}, \bibinfo
  {author} {\bibfnamefont {K.}~\bibnamefont {Shibata}}, \bibinfo {author}
  {\bibfnamefont {Y.}~\bibnamefont {Tokunaga}}, \bibinfo {author}
  {\bibfnamefont {Y.}~\bibnamefont {Taguchi}}, \ and\ \bibinfo {author}
  {\bibfnamefont {Y.}~\bibnamefont {Tokura}},\ }\href {\doibase
  10.1103/PhysRevB.91.054411} {\bibfield  {journal} {\bibinfo  {journal} {Phys.
  Rev. B}\ }\textbf {\bibinfo {volume} {91}},\ \bibinfo {pages} {054411}
  (\bibinfo {year} {2015})}\BibitemShut {NoStop}%
\bibitem [{\citenamefont {Leonov}\ \emph {et~al.}(2016)\citenamefont {Leonov},
  \citenamefont {Togawa}, \citenamefont {Monchesky}, \citenamefont {Bogdanov},
  \citenamefont {Kishine}, \citenamefont {Kousaka}, \citenamefont {Miyagawa},
  \citenamefont {Koyama}, \citenamefont {Akimitsu}, \citenamefont {Koyama},
  \citenamefont {Harada}, \citenamefont {Mori}, \citenamefont {McGrouther},
  \citenamefont {Lamb}, \citenamefont {Krajnak}, \citenamefont {McVitie},
  \citenamefont {Stamps},\ and\ \citenamefont {Inoue}}]{Leonov:2016prl}%
  \BibitemOpen
  \bibfield  {author} {\bibinfo {author} {\bibfnamefont {A.~O.}\ \bibnamefont
  {Leonov}}, \bibinfo {author} {\bibfnamefont {Y.}~\bibnamefont {Togawa}},
  \bibinfo {author} {\bibfnamefont {T.~L.}\ \bibnamefont {Monchesky}}, \bibinfo
  {author} {\bibfnamefont {A.~N.}\ \bibnamefont {Bogdanov}}, \bibinfo {author}
  {\bibfnamefont {J.}~\bibnamefont {Kishine}}, \bibinfo {author} {\bibfnamefont
  {Y.}~\bibnamefont {Kousaka}}, \bibinfo {author} {\bibfnamefont
  {M.}~\bibnamefont {Miyagawa}}, \bibinfo {author} {\bibfnamefont
  {T.}~\bibnamefont {Koyama}}, \bibinfo {author} {\bibfnamefont
  {J.}~\bibnamefont {Akimitsu}}, \bibinfo {author} {\bibfnamefont
  {T.}~\bibnamefont {Koyama}}, \bibinfo {author} {\bibfnamefont
  {K.}~\bibnamefont {Harada}}, \bibinfo {author} {\bibfnamefont
  {S.}~\bibnamefont {Mori}}, \bibinfo {author} {\bibfnamefont {D.}~\bibnamefont
  {McGrouther}}, \bibinfo {author} {\bibfnamefont {R.}~\bibnamefont {Lamb}},
  \bibinfo {author} {\bibfnamefont {M.}~\bibnamefont {Krajnak}}, \bibinfo
  {author} {\bibfnamefont {S.}~\bibnamefont {McVitie}}, \bibinfo {author}
  {\bibfnamefont {R.~L.}\ \bibnamefont {Stamps}}, \ and\ \bibinfo {author}
  {\bibfnamefont {K.}~\bibnamefont {Inoue}},\ }\href {\doibase
  10.1103/PhysRevLett.117.087202} {\bibfield  {journal} {\bibinfo  {journal}
  {Phys. Rev. Lett.}\ }\textbf {\bibinfo {volume} {117}},\ \bibinfo {pages}
  {087202} (\bibinfo {year} {2016})}\BibitemShut {NoStop}%
\bibitem [{\citenamefont {Wilson}\ \emph {et~al.}(2012)\citenamefont {Wilson},
  \citenamefont {Karhu}, \citenamefont {Quigley}, \citenamefont {R\"o\ss{}ler},
  \citenamefont {Butenko}, \citenamefont {Bogdanov}, \citenamefont
  {Robertson},\ and\ \citenamefont {Monchesky}}]{Wilson:2012prb}%
  \BibitemOpen
  \bibfield  {author} {\bibinfo {author} {\bibfnamefont {M.~N.}\ \bibnamefont
  {Wilson}}, \bibinfo {author} {\bibfnamefont {E.~A.}\ \bibnamefont {Karhu}},
  \bibinfo {author} {\bibfnamefont {A.~S.}\ \bibnamefont {Quigley}}, \bibinfo
  {author} {\bibfnamefont {U.~K.}\ \bibnamefont {R\"o\ss{}ler}}, \bibinfo
  {author} {\bibfnamefont {A.~B.}\ \bibnamefont {Butenko}}, \bibinfo {author}
  {\bibfnamefont {A.~N.}\ \bibnamefont {Bogdanov}}, \bibinfo {author}
  {\bibfnamefont {M.~D.}\ \bibnamefont {Robertson}}, \ and\ \bibinfo {author}
  {\bibfnamefont {T.~L.}\ \bibnamefont {Monchesky}},\ }\href {\doibase
  10.1103/PhysRevB.86.144420} {\bibfield  {journal} {\bibinfo  {journal} {Phys.
  Rev. B}\ }\textbf {\bibinfo {volume} {86}},\ \bibinfo {pages} {144420}
  (\bibinfo {year} {2012})}\BibitemShut {NoStop}%
\bibitem [{\citenamefont {Wilson}\ \emph {et~al.}(2013)\citenamefont {Wilson},
  \citenamefont {Karhu}, \citenamefont {Lake}, \citenamefont {Quigley},
  \citenamefont {Meynell}, \citenamefont {Bogdanov}, \citenamefont {Fritzsche},
  \citenamefont {R\"o\ss{}ler},\ and\ \citenamefont
  {Monchesky}}]{Wilson:2013prb}%
  \BibitemOpen
  \bibfield  {author} {\bibinfo {author} {\bibfnamefont {M.~N.}\ \bibnamefont
  {Wilson}}, \bibinfo {author} {\bibfnamefont {E.~A.}\ \bibnamefont {Karhu}},
  \bibinfo {author} {\bibfnamefont {D.~P.}\ \bibnamefont {Lake}}, \bibinfo
  {author} {\bibfnamefont {A.~S.}\ \bibnamefont {Quigley}}, \bibinfo {author}
  {\bibfnamefont {S.}~\bibnamefont {Meynell}}, \bibinfo {author} {\bibfnamefont
  {A.~N.}\ \bibnamefont {Bogdanov}}, \bibinfo {author} {\bibfnamefont
  {H.}~\bibnamefont {Fritzsche}}, \bibinfo {author} {\bibfnamefont {U.~K.}\
  \bibnamefont {R\"o\ss{}ler}}, \ and\ \bibinfo {author} {\bibfnamefont
  {T.~L.}\ \bibnamefont {Monchesky}},\ }\href {\doibase
  10.1103/PhysRevB.88.214420} {\bibfield  {journal} {\bibinfo  {journal} {Phys.
  Rev. B}\ }\textbf {\bibinfo {volume} {88}},\ \bibinfo {pages} {214420}
  (\bibinfo {year} {2013})}\BibitemShut {NoStop}%
\bibitem [{\citenamefont {Yokouchi}\ \emph {et~al.}(2015)\citenamefont
  {Yokouchi}, \citenamefont {Kanazawa}, \citenamefont {Tsukazaki},
  \citenamefont {Kozuka}, \citenamefont {Kikkawa}, \citenamefont {Taguchi},
  \citenamefont {Kawasaki}, \citenamefont {Ichikawa}, \citenamefont {Kagawa},\
  and\ \citenamefont {Tokura}}]{Tomoyuki:2015jpsj}%
  \BibitemOpen
  \bibfield  {author} {\bibinfo {author} {\bibfnamefont {T.}~\bibnamefont
  {Yokouchi}}, \bibinfo {author} {\bibfnamefont {N.}~\bibnamefont {Kanazawa}},
  \bibinfo {author} {\bibfnamefont {A.}~\bibnamefont {Tsukazaki}}, \bibinfo
  {author} {\bibfnamefont {Y.}~\bibnamefont {Kozuka}}, \bibinfo {author}
  {\bibfnamefont {A.}~\bibnamefont {Kikkawa}}, \bibinfo {author} {\bibfnamefont
  {Y.}~\bibnamefont {Taguchi}}, \bibinfo {author} {\bibfnamefont
  {M.}~\bibnamefont {Kawasaki}}, \bibinfo {author} {\bibfnamefont
  {M.}~\bibnamefont {Ichikawa}}, \bibinfo {author} {\bibfnamefont
  {F.}~\bibnamefont {Kagawa}}, \ and\ \bibinfo {author} {\bibfnamefont
  {Y.}~\bibnamefont {Tokura}},\ }\href {\doibase 10.7566/JPSJ.84.104708}
  {\bibfield  {journal} {\bibinfo  {journal} {J. Phys. Soc. Jpn.}\ }\textbf
  {\bibinfo {volume} {84}},\ \bibinfo {pages} {104708} (\bibinfo {year}
  {2015})}\BibitemShut {NoStop}%
\bibitem [{\citenamefont {Karhu}\ \emph {et~al.}(2010)\citenamefont {Karhu},
  \citenamefont {Kahwaji}, \citenamefont {Monchesky}, \citenamefont {Parsons},
  \citenamefont {Robertson},\ and\ \citenamefont {Maunders}}]{Karhu:2010prb}%
  \BibitemOpen
  \bibfield  {author} {\bibinfo {author} {\bibfnamefont {E.}~\bibnamefont
  {Karhu}}, \bibinfo {author} {\bibfnamefont {S.}~\bibnamefont {Kahwaji}},
  \bibinfo {author} {\bibfnamefont {T.~L.}\ \bibnamefont {Monchesky}}, \bibinfo
  {author} {\bibfnamefont {C.}~\bibnamefont {Parsons}}, \bibinfo {author}
  {\bibfnamefont {M.~D.}\ \bibnamefont {Robertson}}, \ and\ \bibinfo {author}
  {\bibfnamefont {C.}~\bibnamefont {Maunders}},\ }\href {\doibase DOI
  10.1103/PhysRevB.82.184417} {\bibfield  {journal} {\bibinfo  {journal} {Phys.
  Rev. B}\ }\textbf {\bibinfo {volume} {82}},\ \bibinfo {pages} {184417}
  (\bibinfo {year} {2010})}\BibitemShut {NoStop}%
\bibitem [{\citenamefont {Karhu}\ \emph {et~al.}(2011)\citenamefont {Karhu},
  \citenamefont {Kahwaji}, \citenamefont {Robertson}, \citenamefont
  {Fritzsche}, \citenamefont {Kirby}, \citenamefont {Majkrzak},\ and\
  \citenamefont {Monchesky}}]{Karhu:2011prb}%
  \BibitemOpen
  \bibfield  {author} {\bibinfo {author} {\bibfnamefont {E.~A.}\ \bibnamefont
  {Karhu}}, \bibinfo {author} {\bibfnamefont {S.}~\bibnamefont {Kahwaji}},
  \bibinfo {author} {\bibfnamefont {M.~D.}\ \bibnamefont {Robertson}}, \bibinfo
  {author} {\bibfnamefont {H.}~\bibnamefont {Fritzsche}}, \bibinfo {author}
  {\bibfnamefont {B.~J.}\ \bibnamefont {Kirby}}, \bibinfo {author}
  {\bibfnamefont {C.~F.}\ \bibnamefont {Majkrzak}}, \ and\ \bibinfo {author}
  {\bibfnamefont {T.~L.}\ \bibnamefont {Monchesky}},\ }\href {\doibase
  10.1103/PhysRevB.84.060404} {\bibfield  {journal} {\bibinfo  {journal} {Phys.
  Rev. B}\ }\textbf {\bibinfo {volume} {84}},\ \bibinfo {pages} {060404}
  (\bibinfo {year} {2011})}\BibitemShut {NoStop}%
\bibitem [{\citenamefont {Karhu}\ \emph {et~al.}(2012)\citenamefont {Karhu},
  \citenamefont {R\"o\ss{}ler}, \citenamefont {Bogdanov}, \citenamefont
  {Kahwaji}, \citenamefont {Kirby}, \citenamefont {Fritzsche}, \citenamefont
  {Robertson}, \citenamefont {Majkrzak},\ and\ \citenamefont
  {Monchesky}}]{Karhu:2012prb}%
  \BibitemOpen
  \bibfield  {author} {\bibinfo {author} {\bibfnamefont {E.~A.}\ \bibnamefont
  {Karhu}}, \bibinfo {author} {\bibfnamefont {U.~K.}\ \bibnamefont
  {R\"o\ss{}ler}}, \bibinfo {author} {\bibfnamefont {A.~N.}\ \bibnamefont
  {Bogdanov}}, \bibinfo {author} {\bibfnamefont {S.}~\bibnamefont {Kahwaji}},
  \bibinfo {author} {\bibfnamefont {B.~J.}\ \bibnamefont {Kirby}}, \bibinfo
  {author} {\bibfnamefont {H.}~\bibnamefont {Fritzsche}}, \bibinfo {author}
  {\bibfnamefont {M.~D.}\ \bibnamefont {Robertson}}, \bibinfo {author}
  {\bibfnamefont {C.~F.}\ \bibnamefont {Majkrzak}}, \ and\ \bibinfo {author}
  {\bibfnamefont {T.~L.}\ \bibnamefont {Monchesky}},\ }\href {\doibase
  10.1103/PhysRevB.85.094429} {\bibfield  {journal} {\bibinfo  {journal} {Phys.
  Rev. B}\ }\textbf {\bibinfo {volume} {85}},\ \bibinfo {pages} {094429}
  (\bibinfo {year} {2012})}\BibitemShut {NoStop}%
\bibitem [{\citenamefont {Huang}\ and\ \citenamefont
  {Chien}(2012)}]{Huang:2012prl}%
  \BibitemOpen
  \bibfield  {author} {\bibinfo {author} {\bibfnamefont {S.~X.}\ \bibnamefont
  {Huang}}\ and\ \bibinfo {author} {\bibfnamefont {C.~L.}\ \bibnamefont
  {Chien}},\ }\href {\doibase 10.1103/PhysRevLett.108.267201} {\bibfield
  {journal} {\bibinfo  {journal} {Phys. Rev. Lett.}\ }\textbf {\bibinfo
  {volume} {108}},\ \bibinfo {pages} {267201} (\bibinfo {year}
  {2012})}\BibitemShut {NoStop}%
\bibitem [{\citenamefont {Porter}\ \emph {et~al.}(2012)\citenamefont {Porter},
  \citenamefont {Creeth},\ and\ \citenamefont {Marrows}}]{Porter:2012prb}%
  \BibitemOpen
  \bibfield  {author} {\bibinfo {author} {\bibfnamefont {N.~A.}\ \bibnamefont
  {Porter}}, \bibinfo {author} {\bibfnamefont {G.~L.}\ \bibnamefont {Creeth}},
  \ and\ \bibinfo {author} {\bibfnamefont {C.~H.}\ \bibnamefont {Marrows}},\
  }\href {\doibase 10.1103/PhysRevB.86.064423} {\bibfield  {journal} {\bibinfo
  {journal} {Phys. Rev. B}\ }\textbf {\bibinfo {volume} {86}},\ \bibinfo
  {pages} {064423} (\bibinfo {year} {2012})}\BibitemShut {NoStop}%
\bibitem [{\citenamefont {Sinha}\ \emph {et~al.}(2014)\citenamefont {Sinha},
  \citenamefont {Porter},\ and\ \citenamefont {Marrows}}]{Sinha:2014prb}%
  \BibitemOpen
  \bibfield  {author} {\bibinfo {author} {\bibfnamefont {P.}~\bibnamefont
  {Sinha}}, \bibinfo {author} {\bibfnamefont {N.~A.}\ \bibnamefont {Porter}}, \
  and\ \bibinfo {author} {\bibfnamefont {C.~H.}\ \bibnamefont {Marrows}},\
  }\href {\doibase 10.1103/PhysRevB.89.134426} {\bibfield  {journal} {\bibinfo
  {journal} {Phys. Rev. B}\ }\textbf {\bibinfo {volume} {89}},\ \bibinfo
  {pages} {134426} (\bibinfo {year} {2014})}\BibitemShut {NoStop}%
\bibitem [{\citenamefont {Engelke}\ \emph {et~al.}(2012)\citenamefont
  {Engelke}, \citenamefont {Reimann}, \citenamefont {Hoffmann}, \citenamefont
  {Gass}, \citenamefont {Menzel},\ and\ \citenamefont
  {S\"{u}llow}}]{Engelke:2012jpsp}%
  \BibitemOpen
  \bibfield  {author} {\bibinfo {author} {\bibfnamefont {J.}~\bibnamefont
  {Engelke}}, \bibinfo {author} {\bibfnamefont {T.}~\bibnamefont {Reimann}},
  \bibinfo {author} {\bibfnamefont {L.}~\bibnamefont {Hoffmann}}, \bibinfo
  {author} {\bibfnamefont {S.}~\bibnamefont {Gass}}, \bibinfo {author}
  {\bibfnamefont {D.}~\bibnamefont {Menzel}}, \ and\ \bibinfo {author}
  {\bibfnamefont {S.}~\bibnamefont {S\"{u}llow}},\ }\href {\doibase
  10.1143/JPSJ.81.124709} {\bibfield  {journal} {\bibinfo  {journal} {J. Phys.
  Soc. Jpn.}\ }\textbf {\bibinfo {volume} {81}},\ \bibinfo {pages} {124709}
  (\bibinfo {year} {2012})}\BibitemShut {NoStop}%
\bibitem [{\citenamefont {Wilson}\ \emph {et~al.}(2014)\citenamefont {Wilson},
  \citenamefont {Butenko}, \citenamefont {Bogdanov},\ and\ \citenamefont
  {Monchesky}}]{Wilson:2014prb}%
  \BibitemOpen
  \bibfield  {author} {\bibinfo {author} {\bibfnamefont {M.~N.}\ \bibnamefont
  {Wilson}}, \bibinfo {author} {\bibfnamefont {A.~B.}\ \bibnamefont {Butenko}},
  \bibinfo {author} {\bibfnamefont {A.~N.}\ \bibnamefont {Bogdanov}}, \ and\
  \bibinfo {author} {\bibfnamefont {T.~L.}\ \bibnamefont {Monchesky}},\ }\href
  {\doibase 10.1103/PhysRevB.89.094411} {\bibfield  {journal} {\bibinfo
  {journal} {Phys. Rev. B}\ }\textbf {\bibinfo {volume} {89}},\ \bibinfo
  {pages} {094411} (\bibinfo {year} {2014})}\BibitemShut {NoStop}%
\bibitem [{\citenamefont {Lancaster}\ \emph {et~al.}(2016)\citenamefont
  {Lancaster}, \citenamefont {Xiao}, \citenamefont {Salman}, \citenamefont
  {Thomas}, \citenamefont {Blundell}, \citenamefont {Pratt}, \citenamefont
  {Clark}, \citenamefont {Prokscha}, \citenamefont {Suter}, \citenamefont
  {Zhang}, \citenamefont {Baker},\ and\ \citenamefont
  {Hesjedal}}]{Lancaster:2016prb}%
  \BibitemOpen
  \bibfield  {author} {\bibinfo {author} {\bibfnamefont {T.}~\bibnamefont
  {Lancaster}}, \bibinfo {author} {\bibfnamefont {F.}~\bibnamefont {Xiao}},
  \bibinfo {author} {\bibfnamefont {Z.}~\bibnamefont {Salman}}, \bibinfo
  {author} {\bibfnamefont {I.~O.}\ \bibnamefont {Thomas}}, \bibinfo {author}
  {\bibfnamefont {S.~J.}\ \bibnamefont {Blundell}}, \bibinfo {author}
  {\bibfnamefont {F.~L.}\ \bibnamefont {Pratt}}, \bibinfo {author}
  {\bibfnamefont {S.~J.}\ \bibnamefont {Clark}}, \bibinfo {author}
  {\bibfnamefont {T.}~\bibnamefont {Prokscha}}, \bibinfo {author}
  {\bibfnamefont {A.}~\bibnamefont {Suter}}, \bibinfo {author} {\bibfnamefont
  {S.~L.}\ \bibnamefont {Zhang}}, \bibinfo {author} {\bibfnamefont {A.~A.}\
  \bibnamefont {Baker}}, \ and\ \bibinfo {author} {\bibfnamefont
  {T.}~\bibnamefont {Hesjedal}},\ }\href {\doibase 10.1103/PhysRevB.93.140412}
  {\bibfield  {journal} {\bibinfo  {journal} {Phys. Rev. B}\ }\textbf {\bibinfo
  {volume} {93}},\ \bibinfo {pages} {140412} (\bibinfo {year}
  {2016})}\BibitemShut {NoStop}%
\bibitem [{\citenamefont {Li}\ \emph {et~al.}(2013)\citenamefont {Li},
  \citenamefont {Kanazawa}, \citenamefont {Yu}, \citenamefont {Tsukazaki},
  \citenamefont {Kawasaki}, \citenamefont {Ichikawa}, \citenamefont {Jin},
  \citenamefont {Kagawa},\ and\ \citenamefont {Tokura}}]{Li:2013prl}%
  \BibitemOpen
  \bibfield  {author} {\bibinfo {author} {\bibfnamefont {Y.}~\bibnamefont
  {Li}}, \bibinfo {author} {\bibfnamefont {N.}~\bibnamefont {Kanazawa}},
  \bibinfo {author} {\bibfnamefont {X.~Z.}\ \bibnamefont {Yu}}, \bibinfo
  {author} {\bibfnamefont {A.}~\bibnamefont {Tsukazaki}}, \bibinfo {author}
  {\bibfnamefont {M.}~\bibnamefont {Kawasaki}}, \bibinfo {author}
  {\bibfnamefont {M.}~\bibnamefont {Ichikawa}}, \bibinfo {author}
  {\bibfnamefont {X.~F.}\ \bibnamefont {Jin}}, \bibinfo {author} {\bibfnamefont
  {F.}~\bibnamefont {Kagawa}}, \ and\ \bibinfo {author} {\bibfnamefont
  {Y.}~\bibnamefont {Tokura}},\ }\href {\doibase
  10.1103/PhysRevLett.110.117202} {\bibfield  {journal} {\bibinfo  {journal}
  {Phys. Rev. Lett.}\ }\textbf {\bibinfo {volume} {110}},\ \bibinfo {pages}
  {117202} (\bibinfo {year} {2013})}\BibitemShut {NoStop}%
\bibitem [{\citenamefont {Monchesky}\ \emph {et~al.}(2014)\citenamefont
  {Monchesky}, \citenamefont {Loudon}, \citenamefont {Robertson},\ and\
  \citenamefont {Bogdanov}}]{Monchesky:2014prl}%
  \BibitemOpen
  \bibfield  {author} {\bibinfo {author} {\bibfnamefont {T.~L.}\ \bibnamefont
  {Monchesky}}, \bibinfo {author} {\bibfnamefont {J.~C.}\ \bibnamefont
  {Loudon}}, \bibinfo {author} {\bibfnamefont {M.~D.}\ \bibnamefont
  {Robertson}}, \ and\ \bibinfo {author} {\bibfnamefont {A.~N.}\ \bibnamefont
  {Bogdanov}},\ }\href {\doibase 10.1103/PhysRevLett.112.059701} {\bibfield
  {journal} {\bibinfo  {journal} {Phys. Rev. Lett.}\ }\textbf {\bibinfo
  {volume} {112}},\ \bibinfo {pages} {059701} (\bibinfo {year}
  {2014})}\BibitemShut {NoStop}%
\bibitem [{\citenamefont {Meynell}\ \emph
  {et~al.}(2014{\natexlab{a}})\citenamefont {Meynell}, \citenamefont {Wilson},
  \citenamefont {Loudon}, \citenamefont {Spitzig}, \citenamefont {Rybakov},
  \citenamefont {Johnson},\ and\ \citenamefont {Monchesky}}]{Meynell:2014prb2}%
  \BibitemOpen
  \bibfield  {author} {\bibinfo {author} {\bibfnamefont {S.~A.}\ \bibnamefont
  {Meynell}}, \bibinfo {author} {\bibfnamefont {M.~N.}\ \bibnamefont {Wilson}},
  \bibinfo {author} {\bibfnamefont {J.~C.}\ \bibnamefont {Loudon}}, \bibinfo
  {author} {\bibfnamefont {A.}~\bibnamefont {Spitzig}}, \bibinfo {author}
  {\bibfnamefont {F.~N.}\ \bibnamefont {Rybakov}}, \bibinfo {author}
  {\bibfnamefont {M.~B.}\ \bibnamefont {Johnson}}, \ and\ \bibinfo {author}
  {\bibfnamefont {T.~L.}\ \bibnamefont {Monchesky}},\ }\href {\doibase
  10.1103/PhysRevB.90.224419} {\bibfield  {journal} {\bibinfo  {journal} {Phys.
  Rev. B}\ }\textbf {\bibinfo {volume} {90}},\ \bibinfo {pages} {224419}
  (\bibinfo {year} {2014}{\natexlab{a}})}\BibitemShut {NoStop}%
\bibitem [{\citenamefont {Bak}\ and\ \citenamefont
  {Jensen}(1980)}]{Bak:1980jpc}%
  \BibitemOpen
  \bibfield  {author} {\bibinfo {author} {\bibfnamefont {P.}~\bibnamefont
  {Bak}}\ and\ \bibinfo {author} {\bibfnamefont {M.~H.}\ \bibnamefont
  {Jensen}},\ }\href {http://dx.doi.org/} {\bibfield  {journal} {\bibinfo
  {journal} {J. Phys. C: Solid State}\ }\textbf {\bibinfo {volume} {13}},\
  \bibinfo {pages} {L881} (\bibinfo {year} {1980})}\BibitemShut {NoStop}%
\bibitem [{\citenamefont {Vansteenkiste}\ \emph {et~al.}(2014)\citenamefont
  {Vansteenkiste}, \citenamefont {Leliaert}, \citenamefont {Dvornik},
  \citenamefont {Helsen}, \citenamefont {Garcia-Sanchez},\ and\ \citenamefont
  {Van~Waeyenberge}}]{Vansteenkiste:2014a}%
  \BibitemOpen
  \bibfield  {author} {\bibinfo {author} {\bibfnamefont {A.}~\bibnamefont
  {Vansteenkiste}}, \bibinfo {author} {\bibfnamefont {J.}~\bibnamefont
  {Leliaert}}, \bibinfo {author} {\bibfnamefont {M.}~\bibnamefont {Dvornik}},
  \bibinfo {author} {\bibfnamefont {M.}~\bibnamefont {Helsen}}, \bibinfo
  {author} {\bibfnamefont {F.}~\bibnamefont {Garcia-Sanchez}}, \ and\ \bibinfo
  {author} {\bibfnamefont {B.}~\bibnamefont {Van~Waeyenberge}},\ }\href
  {\doibase http://dx.doi.org/10.1063/1.4899186} {\bibfield  {journal}
  {\bibinfo  {journal} {AIP Advances}\ }\textbf {\bibinfo {volume} {4}},\
  \bibinfo {eid} {107133} (\bibinfo {year} {2014})}\BibitemShut {NoStop}%
\bibitem [{\citenamefont {Fritzsche}(2005)}]{Fritzsche:2005rsi}%
  \BibitemOpen
  \bibfield  {author} {\bibinfo {author} {\bibfnamefont {H.}~\bibnamefont
  {Fritzsche}},\ }\href {\doibase http://dx.doi.org/10.1063/1.2130666}
  {\bibfield  {journal} {\bibinfo  {journal} {Rev. Sci. Instrum.}\ }\textbf
  {\bibinfo {volume} {76}},\ \bibinfo {eid} {115104} (\bibinfo {year}
  {2005})}\BibitemShut {NoStop}%
\bibitem [{\citenamefont {Meynell}\ \emph
  {et~al.}(2014{\natexlab{b}})\citenamefont {Meynell}, \citenamefont {Wilson},
  \citenamefont {Fritzsche}, \citenamefont {Bogdanov},\ and\ \citenamefont
  {Monchesky}}]{Meynell:2014prb1}%
  \BibitemOpen
  \bibfield  {author} {\bibinfo {author} {\bibfnamefont {S.~A.}\ \bibnamefont
  {Meynell}}, \bibinfo {author} {\bibfnamefont {M.~N.}\ \bibnamefont {Wilson}},
  \bibinfo {author} {\bibfnamefont {H.}~\bibnamefont {Fritzsche}}, \bibinfo
  {author} {\bibfnamefont {A.~N.}\ \bibnamefont {Bogdanov}}, \ and\ \bibinfo
  {author} {\bibfnamefont {T.~L.}\ \bibnamefont {Monchesky}},\ }\href {\doibase
  10.1103/PhysRevB.90.014406} {\bibfield  {journal} {\bibinfo  {journal} {Phys.
  Rev. B}\ }\textbf {\bibinfo {volume} {90}},\ \bibinfo {pages} {014406}
  (\bibinfo {year} {2014}{\natexlab{b}})}\BibitemShut {NoStop}%
\bibitem [{\citenamefont {Bogdanov}\ and\ \citenamefont
  {Hubert}(1994{\natexlab{b}})}]{Bogdanov:1994pss}%
  \BibitemOpen
  \bibfield  {author} {\bibinfo {author} {\bibfnamefont {A.}~\bibnamefont
  {Bogdanov}}\ and\ \bibinfo {author} {\bibfnamefont {A.}~\bibnamefont
  {Hubert}},\ }\href {\doibase 10.1002/pssb.2221860223} {\bibfield  {journal}
  {\bibinfo  {journal} {Phys. Status Solidi B}\ }\textbf {\bibinfo {volume}
  {186}},\ \bibinfo {pages} {527} (\bibinfo {year}
  {1994}{\natexlab{b}})}\BibitemShut {NoStop}%
\bibitem [{\citenamefont {Milde}\ \emph {et~al.}(2013)\citenamefont {Milde},
  \citenamefont {K{\"o}hler}, \citenamefont {Seidel}, \citenamefont {Eng},
  \citenamefont {Bauer}, \citenamefont {Chacon}, \citenamefont {Kindervater},
  \citenamefont {M{\"u}hlbauer}, \citenamefont {Pfleiderer}, \citenamefont
  {Buhrandt}, \citenamefont {Sch{\"u}tte},\ and\ \citenamefont
  {Rosch}}]{Milde:2013sci}%
  \BibitemOpen
  \bibfield  {author} {\bibinfo {author} {\bibfnamefont {P.}~\bibnamefont
  {Milde}}, \bibinfo {author} {\bibfnamefont {D.}~\bibnamefont {K{\"o}hler}},
  \bibinfo {author} {\bibfnamefont {J.}~\bibnamefont {Seidel}}, \bibinfo
  {author} {\bibfnamefont {L.~M.}\ \bibnamefont {Eng}}, \bibinfo {author}
  {\bibfnamefont {A.}~\bibnamefont {Bauer}}, \bibinfo {author} {\bibfnamefont
  {A.}~\bibnamefont {Chacon}}, \bibinfo {author} {\bibfnamefont
  {J.}~\bibnamefont {Kindervater}}, \bibinfo {author} {\bibfnamefont
  {S.}~\bibnamefont {M{\"u}hlbauer}}, \bibinfo {author} {\bibfnamefont
  {C.}~\bibnamefont {Pfleiderer}}, \bibinfo {author} {\bibfnamefont
  {S.}~\bibnamefont {Buhrandt}}, \bibinfo {author} {\bibfnamefont
  {C.}~\bibnamefont {Sch{\"u}tte}}, \ and\ \bibinfo {author} {\bibfnamefont
  {A.}~\bibnamefont {Rosch}},\ }\href {\doibase 10.1126/science.1234657}
  {\bibfield  {journal} {\bibinfo  {journal} {Science}\ }\textbf {\bibinfo
  {volume} {340}},\ \bibinfo {pages} {1076} (\bibinfo {year}
  {2013})}\BibitemShut {NoStop}%
\bibitem [{\citenamefont {Maleyev}\ \emph {et~al.}(1963)\citenamefont
  {Maleyev}, \citenamefont {Baryakhtar},\ and\ \citenamefont
  {Suris}}]{Maleyev:1963spss}%
  \BibitemOpen
  \bibfield  {author} {\bibinfo {author} {\bibfnamefont {S.~V.}\ \bibnamefont
  {Maleyev}}, \bibinfo {author} {\bibfnamefont {V.~G.}\ \bibnamefont
  {Baryakhtar}}, \ and\ \bibinfo {author} {\bibfnamefont {R.~A.}\ \bibnamefont
  {Suris}},\ }\href@noop {} {\bibfield  {journal} {\bibinfo  {journal} {Sov.
  Phys. Solid State}\ }\textbf {\bibinfo {volume} {4}},\ \bibinfo {pages}
  {2533} (\bibinfo {year} {1963})}\BibitemShut {NoStop}%
\bibitem [{\citenamefont {Blume}(1963)}]{Blume:1963}%
  \BibitemOpen
  \bibfield  {author} {\bibinfo {author} {\bibfnamefont {M.}~\bibnamefont
  {Blume}},\ }\href {\doibase 10.1103/PhysRev.130.1670} {\bibfield  {journal}
  {\bibinfo  {journal} {Phys. Rev.}\ }\textbf {\bibinfo {volume} {130}},\
  \bibinfo {pages} {1670} (\bibinfo {year} {1963})}\BibitemShut {NoStop}%
\bibitem [{\citenamefont {Lindenmeyer}\ and\ \citenamefont
  {Hosemann}(1963)}]{Lindenmeyer:1963jap}%
  \BibitemOpen
  \bibfield  {author} {\bibinfo {author} {\bibfnamefont {P.~H.}\ \bibnamefont
  {Lindenmeyer}}\ and\ \bibinfo {author} {\bibfnamefont {R.}~\bibnamefont
  {Hosemann}},\ }\href {http://aip.scitation.org/doi/abs/10.1063/1.1729086}
  {\bibfield  {journal} {\bibinfo  {journal} {J. Appl. Phys.}\ }\textbf
  {\bibinfo {volume} {34}},\ \bibinfo {pages} {42} (\bibinfo {year}
  {1963})}\BibitemShut {NoStop}%
\bibitem [{\citenamefont {Togawa}\ \emph {et~al.}(2015)\citenamefont {Togawa},
  \citenamefont {Koyama}, \citenamefont {Nishimori}, \citenamefont {Matsumoto},
  \citenamefont {McVitie}, \citenamefont {McGrouther}, \citenamefont {Stamps},
  \citenamefont {Kousaka}, \citenamefont {Akimitsu}, \citenamefont {Nishihara},
  \citenamefont {Inoue}, \citenamefont {Bostrem}, \citenamefont {Sinitsyn},
  \citenamefont {Ovchinnikov},\ and\ \citenamefont {Kishine}}]{Togawa:2015prb}%
  \BibitemOpen
  \bibfield  {author} {\bibinfo {author} {\bibfnamefont {Y.}~\bibnamefont
  {Togawa}}, \bibinfo {author} {\bibfnamefont {T.}~\bibnamefont {Koyama}},
  \bibinfo {author} {\bibfnamefont {Y.}~\bibnamefont {Nishimori}}, \bibinfo
  {author} {\bibfnamefont {Y.}~\bibnamefont {Matsumoto}}, \bibinfo {author}
  {\bibfnamefont {S.}~\bibnamefont {McVitie}}, \bibinfo {author} {\bibfnamefont
  {D.}~\bibnamefont {McGrouther}}, \bibinfo {author} {\bibfnamefont {R.~L.}\
  \bibnamefont {Stamps}}, \bibinfo {author} {\bibfnamefont {Y.}~\bibnamefont
  {Kousaka}}, \bibinfo {author} {\bibfnamefont {J.}~\bibnamefont {Akimitsu}},
  \bibinfo {author} {\bibfnamefont {S.}~\bibnamefont {Nishihara}}, \bibinfo
  {author} {\bibfnamefont {K.}~\bibnamefont {Inoue}}, \bibinfo {author}
  {\bibfnamefont {I.~G.}\ \bibnamefont {Bostrem}}, \bibinfo {author}
  {\bibfnamefont {V.~E.}\ \bibnamefont {Sinitsyn}}, \bibinfo {author}
  {\bibfnamefont {A.~S.}\ \bibnamefont {Ovchinnikov}}, \ and\ \bibinfo {author}
  {\bibfnamefont {J.}~\bibnamefont {Kishine}},\ }\href {\doibase
  10.1103/PhysRevB.92.220412} {\bibfield  {journal} {\bibinfo  {journal} {Phys.
  Rev. B}\ }\textbf {\bibinfo {volume} {92}},\ \bibinfo {pages} {220412}
  (\bibinfo {year} {2015})}\BibitemShut {NoStop}%
\bibitem [{\citenamefont {Kanazawa}\ \emph {et~al.}(2016)\citenamefont
  {Kanazawa}, \citenamefont {White}, \citenamefont {R\o{}nnow}, \citenamefont
  {Dewhurst}, \citenamefont {Fujishiro}, \citenamefont {Tsukazaki},
  \citenamefont {Kozuka}, \citenamefont {Kawasaki}, \citenamefont {Ichikawa},
  \citenamefont {Kagawa},\ and\ \citenamefont {Tokura}}]{Kanazawa:2016prb}%
  \BibitemOpen
  \bibfield  {author} {\bibinfo {author} {\bibfnamefont {N.}~\bibnamefont
  {Kanazawa}}, \bibinfo {author} {\bibfnamefont {J.~S.}\ \bibnamefont {White}},
  \bibinfo {author} {\bibfnamefont {H.~M.}\ \bibnamefont {R\o{}nnow}}, \bibinfo
  {author} {\bibfnamefont {C.~D.}\ \bibnamefont {Dewhurst}}, \bibinfo {author}
  {\bibfnamefont {Y.}~\bibnamefont {Fujishiro}}, \bibinfo {author}
  {\bibfnamefont {A.}~\bibnamefont {Tsukazaki}}, \bibinfo {author}
  {\bibfnamefont {Y.}~\bibnamefont {Kozuka}}, \bibinfo {author} {\bibfnamefont
  {M.}~\bibnamefont {Kawasaki}}, \bibinfo {author} {\bibfnamefont
  {M.}~\bibnamefont {Ichikawa}}, \bibinfo {author} {\bibfnamefont
  {F.}~\bibnamefont {Kagawa}}, \ and\ \bibinfo {author} {\bibfnamefont
  {Y.}~\bibnamefont {Tokura}},\ }\href {\doibase 10.1103/PhysRevB.94.184432}
  {\bibfield  {journal} {\bibinfo  {journal} {Phys. Rev. B}\ }\textbf {\bibinfo
  {volume} {94}},\ \bibinfo {pages} {184432} (\bibinfo {year}
  {2016})}\BibitemShut {NoStop}%
\bibitem [{\citenamefont {Bauer}\ and\ \citenamefont
  {Pfleiderer}(2012)}]{Bauer:2012prb}%
  \BibitemOpen
  \bibfield  {author} {\bibinfo {author} {\bibfnamefont {A.}~\bibnamefont
  {Bauer}}\ and\ \bibinfo {author} {\bibfnamefont {C.}~\bibnamefont
  {Pfleiderer}},\ }\href {\doibase 10.1103/PhysRevB.85.214418} {\bibfield
  {journal} {\bibinfo  {journal} {Phys. Rev. B}\ }\textbf {\bibinfo {volume}
  {85}},\ \bibinfo {pages} {214418} (\bibinfo {year} {2012})}\BibitemShut
  {NoStop}%
\end{thebibliography}
\end{document}